\documentclass{aa}  
\usepackage{booktabs}
\usepackage{ulem}
\usepackage{graphicx}
\usepackage{txfonts}
\usepackage{xcolor}
\usepackage{enumitem}
\usepackage{adjustbox}
\usepackage{tabularx}
\begin{document}

   \title{ASW$^2$DF: Census of the obscured star formation in a galaxy cluster in formation at $z=2.2$}

   \author{Y. H. Zhang\inst{1,2,3,4}
   \and H. Dannerbauer\inst{3,4}
   \and J. M. Pérez-Martínez\inst{3,4}
   \and Y. Koyama\inst{5}
   \and X. Z. Zheng\inst{6,1,2}
   \and C. D'Eugenio\inst{3,4}
   \and B. H. C. Emonts\inst{7}
   \and R. Calvi\inst{8}
   \and Z. Chen\inst{9,10}
   \and K. Daikuhara\inst{11}
   \and C. De Breuck\inst{12}
   \and S. Jin\inst{13,14}
   \and T. Kodama\inst{11}
   \and M. D. Lehnert\inst{15}
   \and A. Naufal\inst{16,17}
   \and R. Shimakawa\inst{18,19}
    }
    
   \institute{Purple Mountain Observatory, Chinese Academy of Sciences, 10 Yuanhua Road, Nanjing, 210023, China \\
   \email{yhzhang@pmo.ac.cn}
   \and School of Astronomy and Space Science, University of Science and Technology of China, Hefei, Anhui 230026, China
   \and Instituto de Astrofísica de Canarias (IAC), E-38205 La Laguna, Tenerife, Spain
   \and Universidad de La Laguna, Dpto. Astrofísica, E-38206 La Laguna, Tenerife, Spain
   \and Subaru Telescope, National Astronomical Observatory of Japan, National Institutes of Natural Sciences (NINS), 650 North A’ohoku Place, Hilo, HI 96720, USA
   \and Tsung-Dao Lee Institute and Key Laboratory for Particle Physics, Astrophysics and Cosmology, Ministry of Education, Shanghai Jiao Tong University, Shanghai, 201210, China
   \and National Radio Astronomy Observatory, 520 Edgemont Road, Charlottesville, VA 22903, USA
   \and INAF–Osservatorio Astronomico di Capodimonte, Salita Moiariello 16, 80131 Napoli, Italy
   \and School of Astronomy and Space Science, Nanjing University, Nanjing 210093, China
   \and Key Laboratory of Modern Astronomy and Astrophysics, Nanjing University, Nanjing 210093, China
   \and Astronomical Institute, Tohoku University, 6-3, Aramaki, Aoba, Sendai, Miyagi, 980-8578, Japan
   \and European Southern Observatory, Karl–Schwarzschild–Straße 2, D-85748 Garching bei München, Germany
   \and Cosmic Dawn Center (DAWN), Denmark
   \and DTU Space, Technical University of Denmark, Elektrovej 327, DK-2800 Kgs. Lyngby, Denmark
   \and Université Lyon 1, ENS de Lyon, CNRS UMR5574, Centre de Recherche Astrophysique de Lyon, F-69230 Saint-Genis-Laval, France
   \and Department of Astronomical Science, The Graduate University for Advanced Studies, 2-21-1 Osawa, Mitaka, Tokyo 181-8588, Japan
   \and National Astronomical Observatory of Japan, 2-21-1 Osawa, Mitaka, Tokyo 181-8588, Japan
   \and Waseda Institute for Advanced Study (WIAS), Waseda University, 1-21-1, Nishi-Waseda, Shinjuku, Tokyo 169-0051, Japan
   \and Center for Data Science, Waseda University, 1-6-1, Nishi-Waseda, Shinjuku, Tokyo 169-0051, Japan
    }
   \date{Received September XX, 2023; accepted March XX, 2023}
 
  \abstract
   {We report the results of the deep and wide Atacama Large Millimeter/submillimeter Array (ALMA) 1.2\,mm mapping of the Spiderweb protocluster at $z=2.16$. The observations were divided into six contiguous fields covering a survey area of 19.3\,arcmin$^2$. With $\sim$13h on-source time, the final maps in the six fields reach the 1$\sigma$ rms noise in a range of $40.3-57.1\,\mu$Jy at a spatial resolution of $0\farcs5-0\farcs9$. By using different source extraction codes and careful visual inspection, we detect 47 ALMA sources at a significance higher than 4$\sigma$. We construct the differential and cumulative number counts down to $\sim0.2$\,mJy after the correction for purity and completeness obtained from Monte Carlo simulations. The ALMA 1.2\,mm number counts of dusty star-forming galaxies (DSFGs) in the Spiderweb protocluster are overall two times that of general fields, some fields/regions showing even higher overdensities (more than a factor of 3). This is consistent with the results from previous studies over a larger scale using single-dish instruments. Comparison of the spatial distributions between different populations indicates that our ALMA sources are likely drawn from the same distribution as CO(1-0) emitters from the COALAS large program, but distinct from that of H$\alpha$ emitters. The cosmic SFR density of the ALMA sources is consistent with previous results (e.g. LABOCA 870\,$\mu$m observations) after accounting for the difference in volume. We show that molecular gas masses estimates from dust measurements are not consistent with the ones derived from CO(1-0) and thus have to be taken with caution. The multiplicity fraction of single-dish DSFGs is higher than that of the field. Moreover, two extreme concentrations of ALMA sources are found on the outskirts of the Spiderweb protocluster, with an excess of more than 12 times that of general fields. 
   These results indicate that the ALMA-detected DSFGs are supplied through gas accretion along filaments and are triggered by intense star formation by accretion shocks before falling into the cluster center. The identified two galaxy groups are likely falling into the protocluster center and will trigger new merger events eventually, as indicated in simulations.
   }

   \maketitle
%

\nolinenumbers

\section{Introduction}
As a result of the inhomogeneous mass assembly in the Universe, the ``cosmic web'' is composed of filaments, sheets, and voids \citep{Bond1996}. Galaxy protoclusters were formed at the densest nodes of the cosmic web at high redshift and evolved into mature galaxy clusters in the local Universe \citep{Overzier2016}. Star formation in protoclusters reached its peak at $z\sim2-3$, with a contribution of $\sim$20\,percent to the cosmic star formation rate (SFR) density at the same epoch \citep{Chiang2017}. Therefore, protoclusters at high redshift provide a unique opportunity to investigate the formation and evolution of different galaxy populations in connection with the large-scale structure assembly.

Submillimeter galaxies (SMGs), first discovered in coarse single-dish (sub)mm surveys, are the brightest population of dusty star-forming galaxies \citep[DSFGs,][]{Casey2014} with at least a few mJy at submillimeter (submm) wavelength \citep[e.g.,][]{Blain2002}, whose star formation is heavily dust-obscured and hardly detected in the optical/near-infrared (NIR) bands. This rare population has been studied for more than two decades since its discovery \citep{Smail1997, Hughes1998}, and their number densities are well constrained at different (sub)mm wavelengths \citep[e.g.,][]{Weiss2009, Oliver2010, Scott2012, Casey2013, Chen2013, Geach2017, Magnelli2019, Simpson2019, Shim2020, Bing2023}. Detailed studies on source properties suggest that DSFGs dominate the cosmic SFR density at cosmic noon \citep{Bourne2017, Koprowski2017, Madau2014}, the peak epoch of cosmic star formation and active galactic nucleus (AGN) activities \citep{Hopkins2006, Zheng2009}, and will eventually evolve into elliptical galaxies in the local Universe \citep[e.g.,][]{Lutz2001, Smail2002, Swinbank2006, Gullberg2019}. 

However, most previous studies of DSFGs are based on observations with single-dish facilities such as JCMT/SCUBA2 \citep{Holland1999, Holland2013}, APEX/LABOCA \citep{Siringo2009} and ASTE/AzTEC \citep{Wilson2008} with a coarse resolution (15$^{\prime\prime}-$30$^{\prime\prime}$), thus hampered by source blending and counterpart identification \citep{Chen2016, An2018}. 
The high confusion noise allows only the detection of bright SMGs \citep[e.g.,][]{Blain2002}. Thanks to the Atacama Large Millimeter/submillimeter Array \citep[ALMA;][]{Wootten2009} with its unprecedented resolution and sensitivity, we have the opportunity to probe DSFGs down to tens of $\mu$Jy at a sub-arcsec resolution, extending to the `normal' star-forming galaxy population with an SFR of tens of solar masses per year \citep{Aravena2016, Aravena2020}. Extensive efforts have been made to investigate the DSFGs with deep ALMA contiguous mapping surveys in general fields \citep[e.g.,][]{Hatsukade2016, Hatsukade2018, Dunlop2017, Gonzalez2017, Gonzalez2019, Gonzalez2020, Franco2018, Decarli2019, Gomez2022}, which helped in our understanding of the dust and gas properties of this population in the distant Universe \citep[see][for a review]{Hodge2020}.

Moreover, SMGs are located preferentially in massive dark-matter halos \citep{Blain2004, Hickox2012, Wilkinson2017} similar to those around radio galaxies \citep[e.g.,][]{Stevens2003, Humphrey2011, Rigby2014, Zeballos2018} and quasars \citep[e.g.,][]{Priddey2008, Stevens2010, Carrera2011, Jones2017, Wethers2020}, thus can serve as a tracer for protoclusters \citep{Calvi2023}. These halos reside in overdense environments and hold intense star formation, making them an ideal laboratory to study where and how extreme starbursts take place. Overdensities of SMGs have been confirmed in protoclusters within a wide redshift range z$\sim2-7$ (\citealp[e.g.,][]{DeBreuck2004, Greve2007, Tamura2009, Dannerbauer2014, Casey2016, AB2018, Wang2021, Zhang2022, Zeng2024, Zhoudz2024}; \citealp[see][for a review]{Alberts2022}). These studies are based on observations with single-dish telescopes and thus can only reveal the overall properties of the brightest SMGs. Therefore, ALMA (contiguous) mapping of these SMG overdensities is indispensable to study DSFGs individually and their connection with the surrounding large-scale structure. Due to a limited number of deep ALMA mapping surveys on such extreme structures \citep[e.g.][]{Umehata2015, Umehata2017, Umehata2018, Zhou2020, Pensabene2024, Wang2024}, it is strongly demanded to create a census of DSFGs in protoclusters to enlarge the sample and investigate the roles that environment plays in forming the DSFG population.

One of the most overdense structures is the prominent Spiderweb protocluster at $z=2.16$, which has been studied for more than twenty years since its discovery \citep{Kurk2000, Pentericci2000}. The high redshift radio galaxy MRC1138-262 \citep{Roettgering1994, Pentericci1997} at the center of the protocluster, has a clumpy morphology with a large number of companion galaxies \citep{Miley2006, Hatch2009} and surrounded by an extended molecular gas reservoir \citep{Emonts2016, Emonts2018}. The overdense environment of the Spiderweb protocluster has been confirmed via various populations, including X-ray detections \citep{Pentericci2002, Croft2005, Tozzi2022a, Tozzi2022b}, Ly$\alpha$ emitters \citep[LAEs;][]{Pentericci2000, Kurk2000}, H$\alpha$ emitters \citep[HAEs;][]{Kuiper2011, Koyama2013, Shimakawa2014, Shimakawa2018b, Daikuhara2024}, extremely red objects \citep[EROs;][]{Kurk2004a}, (sub)millimeter galaxies \citep{Rigby2014, Dannerbauer2014, Zeballos2018}, and CO(1-0) emitters \citep{Jin2021}. The overdensity signature motivates the study of the environmental effects on the member galaxies in terms of UV morphologies \citep{Naufal2023}, gas-phase metallicity \citep{Perez2023}, AGN activity \citep{Tozzi2022a, Shimakawa2024} and the cold molecular gas content \citep{Dannerbauer2017, Emonts2016, Emonts2018, Chen2024}. Observations of hundreds of hours towards this protocluster have been accumulated at a very wide wavelength range from X-ray to radio, making it an ideal target to study DSFGs in such complex environments. Moreover, \citet{Tozzi2022b} reported a diffuse X-ray emission centered on the Spiderweb galaxy over a scale of $\sim100\,$kpc, powered by the inverse-Compton scattering associated with the radio jets. The hot, nascent intracluster medium (ICM) has been detected through the Sunyaev–Zeldovich effect \citep{Mascolo2023}. Especially, the later result shows that this structure will evolve into a mature galaxy cluster in the local Universe.

In this paper, we report the first results of the ALMA Spiderweb Wide and Deep Field (ASW$^2$DF) survey, mapping dusty star-forming galaxies through ALMA band 6 observations on the Spiderweb protocluster at $z=2.16$. In Section~\ref{sec: obs}, we introduce the observations and data reduction, along with a summary of the data products. We describe the source extraction and the constructed source catalog in Section~\ref{sec: catalog}. We present the Monte Carlo simulation and the number count results in Section~\ref{sec: number counts}. In Section~\ref{sec: discussion}, we compare our results with the literature and estimate the overdensity of DSFGs in the Spiderweb protocluster. We summarize our results in Section~\ref{sec: summary}. For simplicity, we refer to sources detected by submm/mm single-dish telescopes to as SMGs, and sources detected by ALMA as DSFGs throughout this paper. We adopt a flat $\Lambda$CDM cosmology with $H_0=71$\,kms\,$^{-1}$\,Mpc$^{-1}$, $\Omega _{\rm \Lambda}$=0.7 \citep{Spergel2003, Spergel2007} throughout the paper, giving a scale of 0.490\,physical\,Mpc (pMpc) or 1.549\,comoving\,Mpc (cMpc)\,per\,arcmin at $z=2.16$.

\begin{figure*}[h!]
\centering
\includegraphics[width=\textwidth]{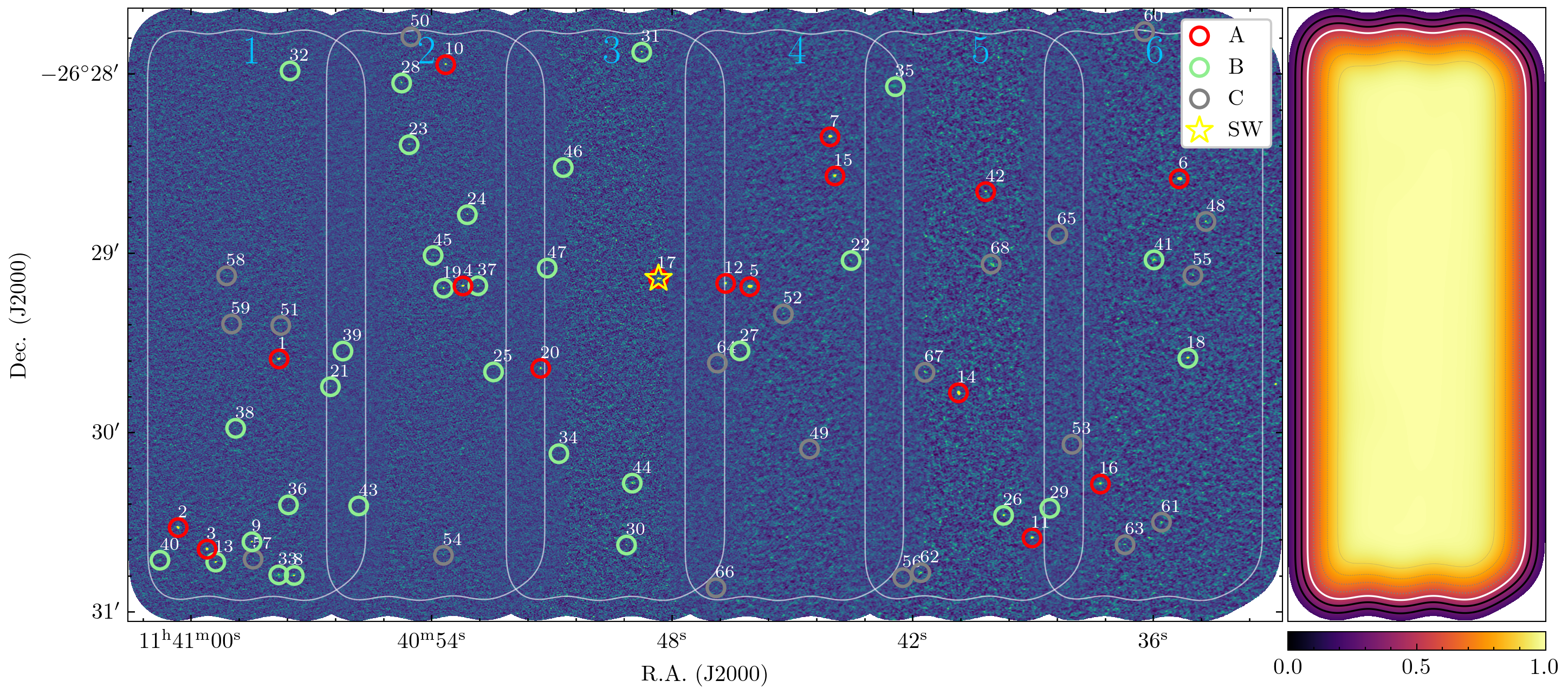}
\caption{{\bf Left:} ALMA 1.2\,mm continuum map of the ASW$^2$DF survey. The fields from left to right are coded by the numbers 1 through 6. 
Our ALMA detections are classified into three categories and marked as red, green, grey circles, in which A for the ``best'' sources identified by all codes and visual inspection, B for the remaining ones in the main catalog, and C for the sources in the supplementary catalog (see Sec. \ref{sec: source extraction}).
The IDs of the ALMA detections in the main and supplementary catalog are marked as white numbers, as listed in Table \ref{table: main catalog} and \ref{table: supp catalog}. 
The white curves enclose the coverage where the primary beam response larger than 0.5 in each field. The images before the primary beam correction are used for clarity. {\bf Right:} The primary beam response of Field 1 shown as an example. We note that the primary beam response in each field is identical, but characterized by various 1$\sigma$ noise. The contours from inner to outer regions are the primary beam correction starting from 0.9 decreasing in a step of 0.1. The white contour surrounds the coverage where PB\textgreater0.5, which will be used for our following analysis.}
     \label{fig: detections}
\end{figure*}

\section{Observations and data reduction}
\label{sec: obs}
ASW$^2$DF is a 1.2 mm galaxy survey aiming to reveal dusty star-forming galaxies in the Spiderweb protocluster. This survey covers the central part of the filamentary structure traced by HAEs \citep{Koyama2013, Shimakawa2018b}, and is complemented by the existing COALAS CO(1-0) large program \citep{Jin2021}. In this section, we describe the details of this survey, as well as the data reduction and the products.

\subsection{Observations}
The ASW$^2$DF mapping observations include six contiguous fields around on the Spiderweb galaxy, each field is made up by a 87-pointing mosaic covering an area of $1^\prime\times3^\prime$. The observations were carried out in Cycle 8 (Project ID: 2021.1.00435.S; PI: Y. Koyama) between 31 December 2021 and 17 April 2022 in ALMA band 6.  Three execution blocks (EBs) have been conducted in each of Field 1, 2, 3, 4, and four EBs in each of Field 5 and Field 6. We note that one EB failed due to antenna issues. The total number of available antennas ranges from 42 to 46. All the EBs in Field 1$-$3 and two in Field 4 were observed in C4 configuration, with a baseline range of 14.9$-$976.6\,m and a theoretical resolution of 0\farcs4. The remaining eight EBs in Field 4$-$6 were observed in C2 configuration, with a baseline range of 15.0\,m to 500.2\,m and a theoretical resolution of 0\farcs7. The maximum recoverable scales are 4\farcs5 and 8\farcs2 in C4 and C2 configurations, respectively. The correlator was set up at a single tuning, with four spectral windows centered at 239.0, 241.0, 255.0 and 257.0\,GHz respectively. Each spectral window contains 128 channels of 15.625\,MHz each in dual polarization with a bandwidth of 1.875\,GHz. The tuning setup also covers the redshifted CO(7-6) and CI(2-1) lines ($\nu_{\rm rest}=806.65$ and 809.34\,GHz) at z$\sim$2.16, providing the possibility of detecting molecular gas in the Spiderweb protocluster. We will report results on these lines in a later publication.

One quasar J1256$-$0547 was used for flux and bandpass calibrations in 11 EBs (Field 1$-$3 and two in Field 4), and another quasar J1037$-$2934 was used in seven EBs (Field 5-6 and one in Field 4). The remaining one EB of Field 4 used quasar J1107$-$4449 as a flux and bandpass calibrator while quasar J1146$-$2859 was observed as the phase calibrator for all EBs. Meanwhile, all the calibrators were used for calibrating the precipitable water vapor (PWV). The average PWV is 0.3-2.4\,mm with a median of 0.9\,mm. The on source time was 2.03\,h for each field (except 2.70\,h for Field 4), resulting in a total on source time of 12.84\,h. 

\begin{figure*}
\sidecaption
\includegraphics[width=6cm]{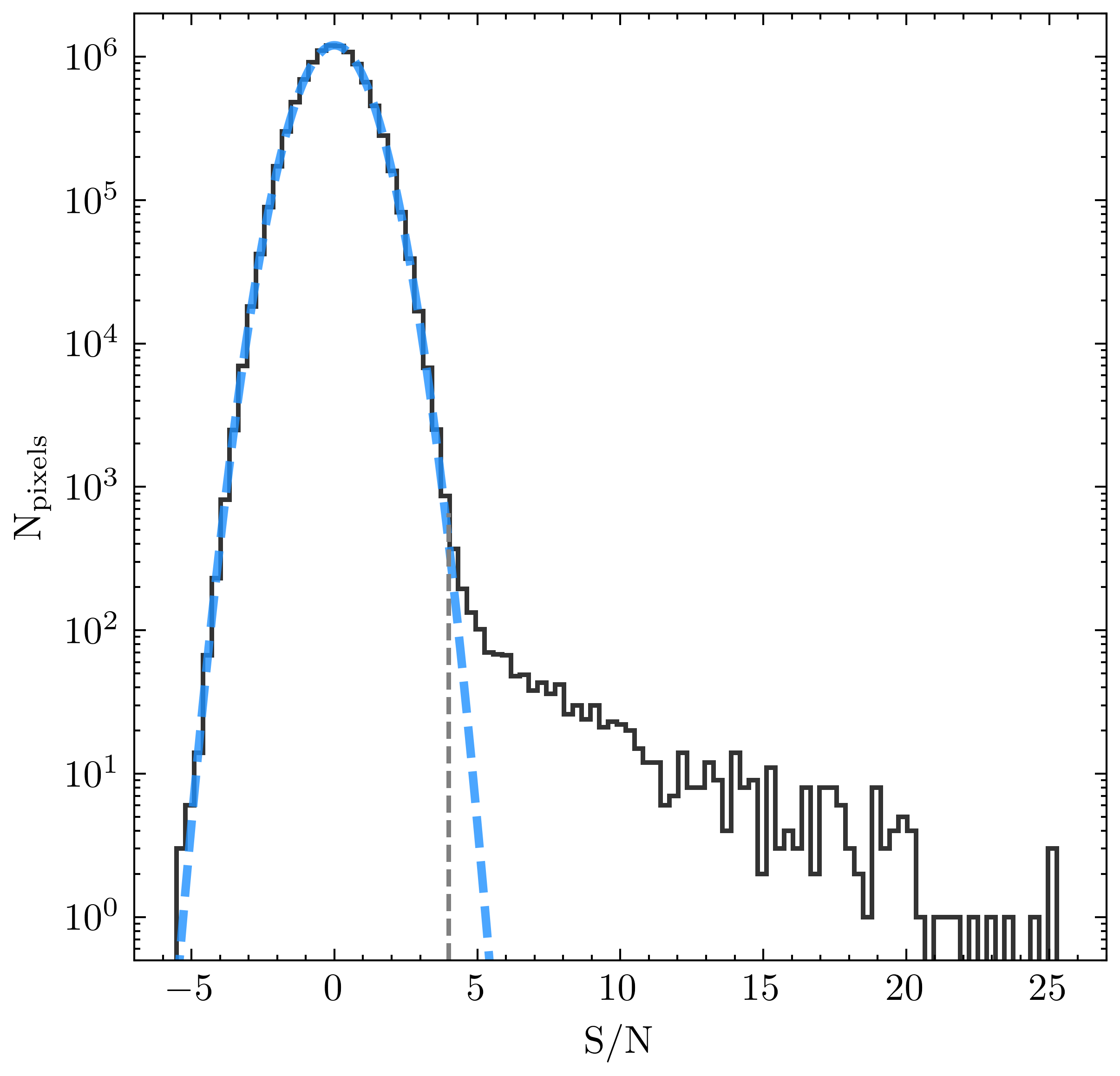}
\includegraphics[width=5.9cm]{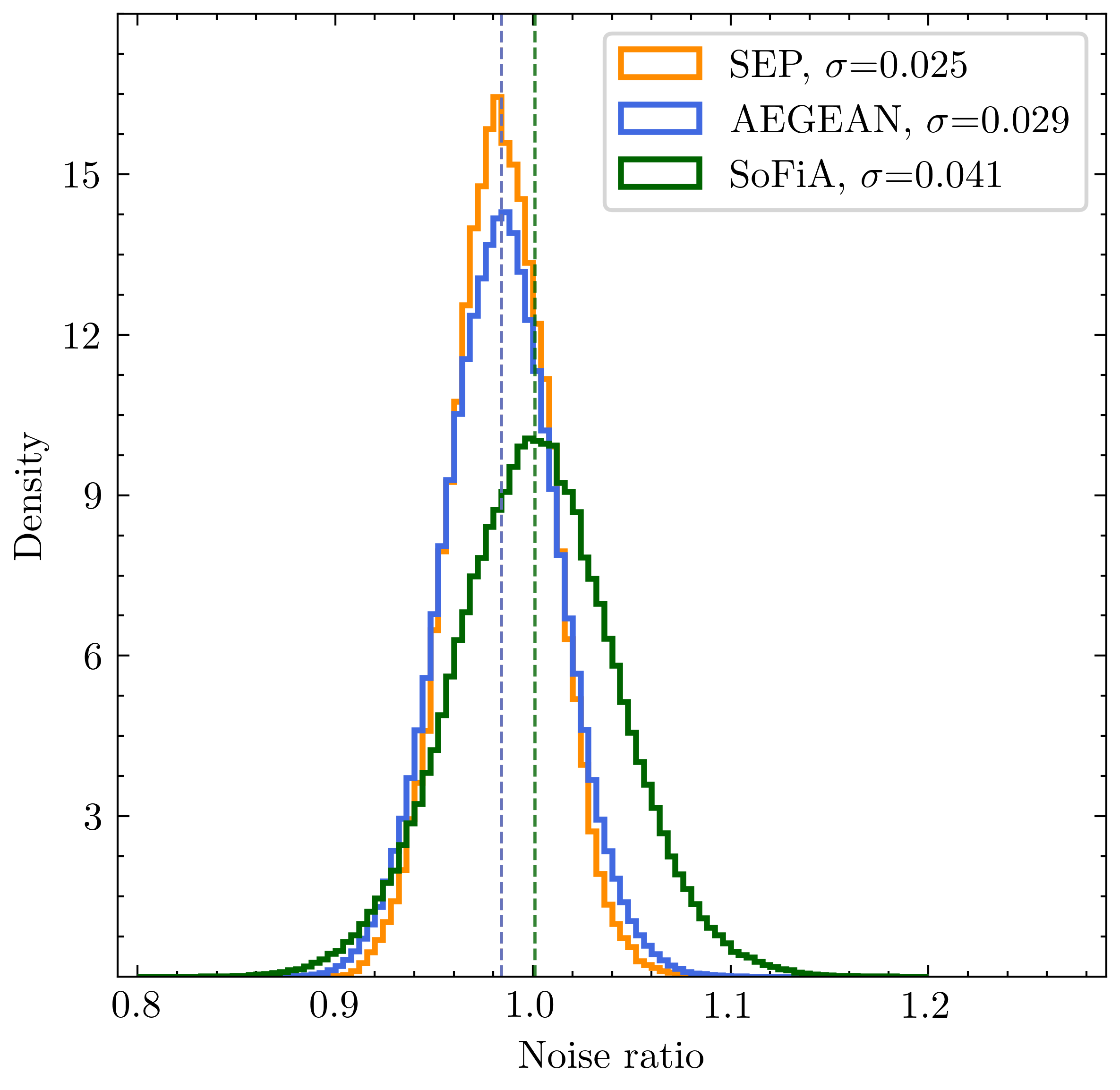}
\caption{{\bf Left:} Pixel distribution of the S/N maps in all six fields. The S/N map is obtained by dividing the science map and the typical 1$\sigma$ noise in each field. The vertical dashed line shows the adopted 4$\sigma$ threshold during the source extraction. {\bf Right:} The distributions of the noise ratios between the noise estimated from the ``individual'' and ``fixed'' methods. The results from SEP, AEGEAN and {\sc SoFiA} are shown in yellow, blue and green. The vertical dashed lines mark the median value from the corresponding ratio distributions. The median noise ratios of SEP and AEGEAN are very close and overlapped, around 0.984. \\ \\}
\label{fig: noise_ratio}
\end{figure*}

\subsection{Data reduction}
The Common Astronomy Software Applications \citep[CASA version 6.2.1.7;][]{CASA2022} with the ALMA pipeline version 2021.2.0.128 \citep{Hunter2023} were used to perform the data reduction. The visibilities were calibrated  and flagged using the scripts provided by the ALMA archive.  The imaging of the visibility data sets were done by using the task {\tt TCLEAN} in CASA, within the ``mosaic'' gridder mode and the ``hogbom'' deconvolver. The ``Briggs'' data weighting with the robust parameter of 0.5 was employed to optimize between angular resolution, noise, and sidelobe levels \citep{Briggs1995}. The multi-frequency synthesis algorithm was used to combine the four spectral windows into the single dirty continuum image. We note that the channels at the edge of the spectral window and the bad baselines, as well as the potential emission lines mentioned above were flagged for the continuum imaging. The final available bandwidth is 6.375\,GHz. After measuring the 1$\sigma$ rms within the entire dirty maps, the cleaning process was repeated until reaching the 3$\sigma$ levels in order to obtain the final clean maps.

Due to the variation of available antennas and the observing conditions, the Execution Fraction (EF) was evaluated to quantify the performance of each EB. The EF is the effective quality of the data from each EB normalized to the expected theoretical value assumed by the ALMA scheduling system. Higher EF means that the EB was observed under good weather conditions and has a better sensitivity. The final images in six fields have different EF (3.62$-$5.05) and a sensitivity range from 40.3\,$\mu$Jy to 57.1\,$\mu$Jy. The sensitivity was measured from the standard deviation of the science image before correcting the flux values at the edge of the mosaic for the primary beam response which will be used for the source extraction and discussed in the next section.   Importantly, Field 4 has the best sensitivity due to an additional EB. The resolution of Field 1$-$3 and Field 4$-$6 are different due to the change of configurations. The average resolution is 0\farcs49\,$\times$\,0\farcs34 with a pixel scale of 0\farcs07 for Field 1$-$3, while the average resolution is  0\farcs87\,$\times$\,0\farcs60 with a pixel scale of 0\farcs14 for Field 4$-$6. In Figure \ref{fig: detections}, the different noise levels and resolutions across the six fields in the mosaic map are clearly seen. We summarize the observing date, number of available antennas, baseline lengths, theoretical resolution, on source time, PWV, EF, synthesized beam, position angle and the 1$\sigma$ noise of each field in Table~\ref{tab: obs}.

\section{Source catalog}
\label{sec: catalog}
One of the main aims of the sub-mm/mm continuum imaging is to create a source catalog. Source extraction is thus an important part of the catalog construction and is performed in different ways in the literature. We extract ALMA sources with the {\sc Python} package SEP \citep{Barbary2016}, which uses the core algorithms of {\sc SExtractor} \citep{Bertin1996}. {\sc SExtractor} is the most commonly used program for detecting the sources in ALMA images at different ALMA bands \citep{Simpson2014, Simpson2015, Aravena2016, Fujimoto2016, Oteo2016, Stach2018, Stach2019, Simpson2020, Klitsch2020, Chen2023a, Fujimoto2023}, as it is able to detect sources with complex structures. We also use AEGEAN \citep{Hancock2012, Hancock2018} which is specifically designed and often used in the radio domain, including ALMA maps \citep[e.g.,][]{Umehata2015, Hatsukade2016, Stach2017, Umehata2017, Umehata2018, Ueda2018, Hatsukade2018}. Additionally, we use the Source Finding Application ({\sc SoFiA}) \citep{Serra2015, Westmeier2021} to extract the ALMA sources as a further check, as it is suitable for searching emission lines from data cubes, but it can also be used for continuum source detection in ALMA maps \citep{Klitsch2018, Hamanowicz2023}. In this section, we describe and compare the results of the rms noise estimation and source extraction from different codes. We note that this is the first time that a (detailed) comparison of different source extraction algorithms applied on ALMA data is done.

\subsection{Noise estimation}
\label{sec: noise}
It is essential to make a fair sensitivity estimation (i.e., rms noise) before source extraction. We use science images without the primary beam (PB) correction for rms noise estimation and source extraction, because the pixel values in these images are quite uniform and perfectly follow a Gaussian profile as shown in the left panel of Figure \ref{fig: noise_ratio}. The PB-corrected images will be used for the flux measurement of detected objects.

In general, two methodologies are commonly employed to estimate the noise of ALMA maps. One method is to calculate the standard deviation within an rms box, as the noise of the central pixel. The box is moving in a step of one pixel across the image, and each pixel will get an individual rms noise which we call here the ``individual'' method. The step size may consist of several pixels, resulting in the estimation of rms noise for pixels located in a grid pattern. Interpolation is subsequently employed to derive the complete noise map. In principle, this method captures the noise fluctuation across the entire image. It is worth noting that the sizes of both the rms box and the moving step vary significantly in the literature. For example, the box with $100\times100$ pixels was used for noise estimation in a step of one pixel \citep{Hatsukade2016, Umehata2017, Umehata2018}. \citet{Franco2018} utilized a box size of 100 pixels but two pixels in each step. \citet{Gomez2022} adopted a step size of two pixels, accompanied by a larger box with $200\times200$ pixels. Moreover, the box size can be scaled by the synthesized beam in order to calculate the noise of the central pixel \citep{Dunlop2017, Stach2017}. The differences in those parameters are also seen in the default values suggested by the source extraction codes previously mentioned.

\begin{table}[h!]
\footnotesize
\caption{Key parameters used in SEP, AEGEAN, {\sc SoFiA}.}
\label{tab: se_params}
\centering
\begin{tabular}{ccc}
\toprule
Code & Key Parameters & Values\\
\midrule
& thresh & 4 \\
& err & 1$\sigma$ noise \\
SEP & minarea & 5 \\
& mask & PB\textgreater0.5 \\
& filter\_kernel & None \\
\midrule
& --seedclip & 4 \\
& --forcerms & 1$\sigma$ noise \\
AEGEAN & --forcebkg & 0 \\
& --island & \\
& --negative & \\
\midrule
& scfind.threshold & 4 \\
& linker.minPixels & 5 \\
{\sc SoFiA} & linker.minSizeXY & 1 \\
& linker.minSizeZ & 1 \\
& scfind.kernelsZ & 0 \\
\bottomrule
\end{tabular}
\end{table}

The other method, i.e. the ``fixed'' method, is to calculate the standard deviation of the whole image without the PB correction as the typical 1$\sigma$ rms noise, and the value is fixed across the entire image. This method is used for ALMA maps in both single pointing and contiguous mapping modes \citep{Gonzalez2017, Zavala2018, Gonzalez2020, Zavala2021, Chen2023a, Fujimoto2023}. Note that in both methods, the sigma-clipping iterations are normally applied before the standard deviation calculation to remove the contribution from the bright sources.

We estimate and compare the rms noise of our science images using the ``individual'' and ``fixed'' methods respectively. All the pixels used for noise estimation are within the coverage where the primary beam response is larger than 50\,percent. The following source extraction will also be restricted within this region. The ``individual'' method is performed using SEP, AEGEAN and {\sc SoFiA} respectively, as these tools can directly create the two-dimensional noise map. For consistency, we use the rms box of $100\times100$ pixels with a step size of one pixel to calculate the standard deviation for the central pixel of the box, instead of using the default values in the three codes. The ``fixed'' method is straightforward by calculating the standard deviation of the entire image in each field as the 1$\sigma$ noise, after five iterations of the 5-sigma-clipping process. We find that {\sc SoFiA} can also produce the 1$\sigma$ noise by disabling the parameter {\tt scaleNoise} and the results are identical to our calculation. 

We divide the noise map from the ``individual'' method by the 1$\sigma$ noise from the ``fixed'' method in each field to obtain the noise ratios for a comparison. The right panel of Fig. \ref{fig: noise_ratio} shows the distributions of the noise ratios of three different codes. The median ratio of the results of SEP and AEGEAN is both $\sim$ 0.984, indicating that the ``individual'' noise map is relatively lower than the ``fixed'' 1$\sigma$ noise. This is consistent with our tests that the noise of the central pixel increases with the box size, and the ``fixed'' method is equivalent to treating the entire image as a single box. Similar results were also reported in the literature \citep[$1-3$\% differences,][]{Franco2018, Zavala2021}. The distribution of ratios for the result of {\sc SoFiA} has a higher median value around one together with a larger scatter. We attribute this to the lack of the sigma-clipping process during the noise estimation in {\sc SoFiA}. Although the median noise obtained using the ``individual'' method in all three codes is very close to the 1$\sigma$ noise from the ``fixed'' method, differences between individual pixels may lead an obvious discrepancy between the source catalogs.

\subsection{Source extraction}
\label{sec: source extraction}

Both the two-dimensional noise maps created by SEP, AEGEAN and {\sc SoFiA} in the ``individual'' method, and the 1$\sigma$ noise value in the ``fixed'' method are used for the source extraction. The sources are detected at \textgreater4$\sigma$ in the PB-uncorrected image in each field with these three codes. An important parameter is how many pixels linked to each other above the detection threshold should be considered as a source. This parameter is characterized by {\tt minarea} with a default value of five in SEP, meaning that a source can only be identified when it has at least five contiguous pixels above 4$\sigma$. The five pixels, much smaller than the synthesized beam size, that meet the 4$\sigma$ threshold must be interconnected with no limitations on their arrangement along the x-axis or y-axis. We adopt this value as it is used by default in SEP and {\sc SExtractor} in most previous studies. Lower values will introduce more spurious sources at the same detection threshold. A similar parameter {\tt linker.minPixels} in {\sc SoFiA} is set to five for consistency. There is no identical parameter in AEGEAN. We cut the original catalog and only include sources with more than five adjacent pixels above 4$\sigma$ into the AEGEAN's catalog. The key parameters with the customized values used in three codes are listed in Table \ref{tab: se_params}.

Based on the noise maps from the ``individual'' method, we detect 54, 48 and 49 sources at \textgreater4$\sigma$ through SEP, AEGEAN and SoFiA respectively. The differences in source numbers come from the discrepancies of the noise maps created by SEP, AEGEAN and {\sc SoFiA} following this method. Understanding the reason why these codes result in different noise maps even with the same rms box size ($100\times100$ pixels) is beyond the scope of this paper; here we only focus on the catalogs they produced and make a conclusion on the final catalog construction. Interestingly, all three codes produced the same catalog containing 47 detections when we use the 1$\sigma$ noise values from the ``fixed'' method. This result confirms the reliability of the codes as they can detect identical sources with the same noise characterization. This explains why obtaining different catalogs with ``individual'' method is mainly due to the use of different noise maps. The higher number of detections from the ``individual'' method is probably due to a lower noise estimation or large dispersion in the noise distribution.

We demonstrate that the results of the source extraction are heavily affected by different noise estimates. Though the ``individual'' method with a defined rms box could reflect the fluctuation of the background noise, it is difficult to find a coherent criterion to calculate the scale of the local noise fluctuation as the resolution and pixel scale vary a lot in different surveys. The 1$\sigma$ noise with the ``fixed'' method is more representative as it is directly derived from the standard deviation of the Gaussian profile drawn in the flux distribution. Additionally, the ``fixed'' method produces $\sim2$\,percent higher noise than the median noise from SEP and AEGEAN with ``individual'' method (see Fig. \ref{fig: noise_ratio} right panel), thus could be considered as a conservative way. 

\begin{figure}[h]
\includegraphics[width=\columnwidth]{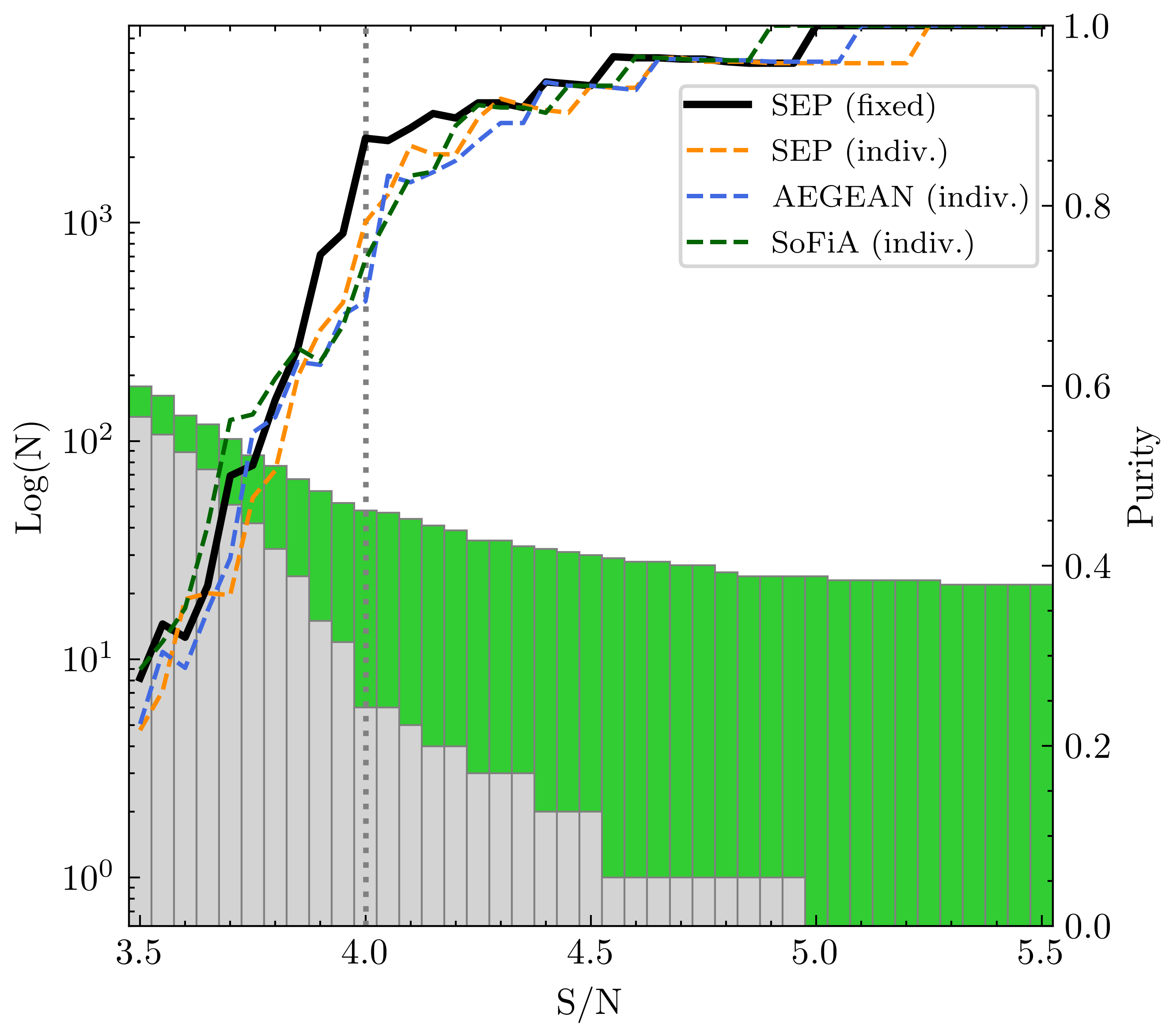}
  \caption{Histograms of positive (green) and negative (grey) detections using the ``fixed'' method at different signal-to-noise ratio thresholds, with the corresponding purity of the positive detections marked as the black curve. The dashed curves show the purity of the ``individual'' method with three different codes (yellow for SEP, blue for AEGEAN, green for {\sc SoFiA}). The vertical dotted line marks the S/N=4 we adopted for the source extraction.}
     \label{fig: purity}
\end{figure}

In this work, we refer to the rms noise value derived from the ``fixed'' method as the typical noise representing the observing depth, which is also the minimum noise of the map after PB correction in each field (Table \ref{tab: obs}). 
Accordingly, there are 47 sources detected by all three codes through the ``fixed'' method. We consider these sources to be primary detections and include them in the main catalog. We note that the number of detections by each code depends on the chosen rms methods. There must be some faint sources that cannot be detected by the ``fixed'' method at 4$\sigma$ significance, but will be detected by the ``individual'' method. We match our main catalog with the results from the ``individual'' method, and find that there are 13 sources detected by at least one of the three codes with the ``individual'' method but cannot be detected by the ``fixed'' method. We mark these 13 sources as tentative sources.

\begin{table*}[ht]
\centering
\caption{Main source catalog.} 
\label{table: main catalog}
\small 
\begin{tabularx}{\textwidth}{ccccccccccc}
\toprule
ID & R.A. & Dec. & S/N$_{\rm p}$ & S/N$_{\rm d}$ & $S_{\rm peak}$ & $S_{\rm fit}$ & $S_{\rm aper}$ & Counterpart & eyes ID  & Cat \\
& (J2000) & (J2000) & & & (mJy) & (mJy) & (mJy) & & & \\
(1) & (2) & (3) & (4) & (5) & (6) & (7) & (8) & (9) & (10) & (11) \\
\midrule
ASW2DF.01 & 11:40:57.80 & -26:29:35.5 & 25.27 & 23.56 & 1.17 ± 0.05 & 1.51 ± 0.05 & 1.81 ± 0.13 & SMG,CO & robust & A \\
ASW2DF.02 & 11:41:00.32 & -26:30:31.9 & 20.31 & 19.02 & 1.04 ± 0.05 & 2.02 ± 0.09 & 2.20 ± 0.11 & SMG,CO & robust & A \\
ASW2DF.03 & 11:40:59.60 & -26:30:39.2 & 20.28 & 19.19 & 0.97 ± 0.05 & 1.78 ± 0.07 & 1.82 ± 0.14 & SMG,CO$_{\rm ext}$ & robust & A \\
ASW2DF.04 & 11:40:53.21 & -26:29:11.1 & 19.69 & 17.93 & 0.92 ± 0.05 & 1.18 ± 0.06 & 1.36 ± 0.12 & SMG & robust & A \\
ASW2DF.05 & 11:40:46.06 & -26:29:11.3 & 18.87 & 17.60 & 0.76 ± 0.04 & 1.27 ± 0.08 & 1.40 ± 0.11 & HAE,CO$_{\rm ext}$ & robust & A \\
ASW2DF.06 & 11:40:35.36 & -26:28:35.1 & 15.28 & 14.27 & 0.68 ± 0.04 & 0.88 ± 0.07 & 0.93 ± 0.11 & — & robust & A \\
ASW2DF.07 & 11:40:44.06 & -26:28:21.1 & 13.61 & 12.21 & 0.56 ± 0.04 & 0.84 ± 0.07 & 0.79 ± 0.10 & — & robust & A \\
ASW2DF.08 & 11:40:57.42 & -26:30:48.0 & 11.78 & 10.93 & 0.66 ± 0.06 & 0.72 ± 0.08 & 1.16 ± 0.14 & — & possible & B \\
ASW2DF.09 & 11:40:58.48 & -26:30:36.7 & 11.31 & 9.93 & 0.53 ± 0.05 & 0.56 ± 0.05 & 0.71 ± 0.10 & — & possible & B \\
ASW2DF.10 & 11:40:53.64 & -26:27:57.0 & 10.32 & 9.52 & 0.54 ± 0.05 & 0.67 ± 0.06 & 0.82 ± 0.14 & SMG & robust & A \\
ASW2DF.11 & 11:40:39.02 & -26:30:35.4 & 10.20 & 9.03 & 0.50 ± 0.05 & 0.59 ± 0.07 & 0.57 ± 0.12 & — & robust & A \\
ASW2DF.12 & 11:40:46.66 & -26:29:10.2 & 10.03 & 8.96 & 0.42 ± 0.04 & 0.57 ± 0.06 & 0.63 ± 0.11 & HAE,CO$_{\rm ext}$ & robust & A \\
ASW2DF.13 & 11:40:59.38 & -26:30:43.5 & 9.99 & 9.51 & 0.51 ± 0.05 & 0.67 ± 0.07 & 0.81 ± 0.15 & — & possible & B \\
ASW2DF.14 & 11:40:40.86 & -26:29:46.8 & 9.93 & 9.31 & 0.48 ± 0.05 & 0.81 ± 0.07 & 0.67 ± 0.10 & CO$_{\rm ext}$ & robust & A \\
ASW2DF.15 & 11:40:43.94 & -26:28:34.2 & 9.39 & 8.63 & 0.38 ± 0.04 & 0.57 ± 0.06 & 0.66 ± 0.11 & — & robust & A \\
ASW2DF.16 & 11:40:37.32 & -26:30:17.3 & 8.36 & 7.76 & 0.38 ± 0.05 & 0.75 ± 0.08 & 0.76 ± 0.12 & HAE,CO & robust & A \\
ASW2DF.17 & 11:40:48.34 & -26:29:08.5 & 8.22 & 7.84 & 0.47 ± 0.06 & 1.00 ± 0.10 & 1.46 ± 0.19 & SMG,CO$_{\rm ext}$ & robust & A \\
ASW2DF.18 & 11:40:35.14 & -26:29:35.2 & 7.07 & 6.67 & 0.32 ± 0.04 & 0.34 ± 0.06 & 0.36 ± 0.12 & — & possible & B \\
ASW2DF.19 & 11:40:53.70 & -26:29:11.8 & 6.90 & 6.56 & 0.32 ± 0.05 & 0.60 ± 0.07 & 0.79 ± 0.11 & — & possible & B \\
ASW2DF.20 & 11:40:51.27 & -26:29:38.6 & 6.57 & 6.05 & 0.40 ± 0.06 & 0.75 ± 0.11 & 0.91 ± 0.18 & HAE,CO & robust & A \\
ASW2DF.21 & 11:40:56.52 & -26:29:44.8 & 6.56 & 5.92 & 0.33 ± 0.05 & 0.29 ± 0.06 & 0.38 ± 0.13 & — & possible & B \\
ASW2DF.22 & 11:40:43.54 & -26:29:02.5 & 6.17 & 5.59 & 0.26 ± 0.04 & 0.29 ± 0.05 & 0.49 ± 0.11 & — & possible & B \\
ASW2DF.23 & 11:40:54.56 & -26:28:23.8 & 5.67 & 5.03 & 0.26 ± 0.05 & 0.47 ± 0.09 & 0.52 ± 0.11 & HAE, CO & possible & B \\
ASW2DF.24 & 11:40:53.10 & -26:28:47.2 & 5.59 & 5.28 & 0.26 ± 0.05 & 0.36 ± 0.06 & 0.28 ± 0.14 & — & possible & B \\
ASW2DF.25 & 11:40:52.45 & -26:29:39.8 & 5.35 & 4.78 & 0.25 ± 0.05 & 0.19 ± 0.05 & 0.18 ± 0.12 & — & possible & B \\
ASW2DF.26 & 11:40:39.73 & -26:30:27.8 & 5.25 & 4.57 & 0.25 ± 0.05 & 0.32 ± 0.07 & 0.32 ± 0.14 & — & possible & B \\
ASW2DF.27 & 11:40:46.31 & -26:29:32.8 & 5.19 & 4.22 & 0.21 ± 0.04 & 0.15 ± 0.04 & 0.17 ± 0.11 & — & possible & B \\
ASW2DF.28 & 11:40:54.74 & -26:28:03.1 & 5.06 & 4.82 & 0.24 ± 0.05 & 0.40 ± 0.08 & 0.56 ± 0.12 & HAE & possible & B \\
ASW2DF.29 & 11:40:38.58 & -26:30:25.5 & 5.05 & 4.55 & 0.38 ± 0.08 & 0.30 ± 0.08 & 0.43 ± 0.19 & CO & marginal & B \\
ASW2DF.30 & 11:40:49.14 & -26:30:37.8 & 4.91 & 4.47 & 0.29 ± 0.06 & 0.23 ± 0.06 & 0.16 ± 0.17 & — & possible & B \\
ASW2DF.31 & 11:40:48.76 & -26:27:52.8 & 4.90 & 4.44 & 0.36 ± 0.07 & 0.65 ± 0.15 & 0.60 ± 0.19 & SMG & possible & B \\
ASW2DF.32 & 11:40:57.52 & -26:27:59.1 & 4.88 & 4.32 & 0.24 ± 0.05 & 0.18 ± 0.05 & 0.13 ± 0.14 & — & — & B \\
ASW2DF.33 & 11:40:57.81 & -26:30:47.8 & 4.83 & 4.66 & 0.28 ± 0.06 & 0.67 ± 0.11 & 1.14 ± 0.15 & SMG, CO$_{\rm ext}$ & cloud-like & B \\
ASW2DF.34 & 11:40:50.82 & -26:30:07.2 & 4.82 & 4.08 & 0.28 ± 0.06 & 0.23 ± 0.05 & 0.26 ± 0.17 & — & — & B \\
ASW2DF.35 & 11:40:42.43 & -26:28:04.3 & 4.81 & 4.14 & 0.26 ± 0.05 & 0.32 ± 0.08 & 0.50 ± 0.13 & — & cloud-like & B \\
ASW2DF.36 & 11:40:57.57 & -26:30:24.3 & 4.74 & 4.21 & 0.22 ± 0.05 & 0.15 ± 0.04 & 0.05 ± 0.09 & — & — & B \\
ASW2DF.37 & 11:40:52.84 & -26:29:11.0 & 4.72 & 4.36 & 0.22 ± 0.05 & 0.35 ± 0.08 & 0.46 ± 0.11 & — & possible & B \\
ASW2DF.38 & 11:40:58.88 & -26:29:58.7 & 4.72 & 4.33 & 0.22 ± 0.05 & 0.16 ± 0.05 & 0.25 ± 0.09 & — & — & B \\
ASW2DF.39 & 11:40:56.21 & -26:29:32.9 & 4.67 & 4.01 & 0.26 ± 0.06 & 0.24 ± 0.06 & 0.12 ± 0.16 & — & possible & B \\
ASW2DF.40 & 11:41:00.77 & -26:30:42.9 & 4.60 & 4.21 & 0.31 ± 0.07 & 0.35 ± 0.08 & 0.56 ± 0.15 & — & possible & B \\
ASW2DF.41 & 11:40:35.98 & -26:29:02.3 & 4.58 & 4.18 & 0.21 ± 0.04 & 0.53 ± 0.10 & 0.65 ± 0.12 & — & cloud-like & B \\
ASW2DF.42 & 11:40:40.18 & -26:28:39.5 & 4.57 & 4.24 & 0.22 ± 0.05 & 0.21 ± 0.06 & 0.06 ± 0.14 & — & robust & A \\
ASW2DF.43 & 11:40:55.81 & -26:30:24.7 & 4.54 & 4.12 & 0.34 ± 0.07 & 0.26 ± 0.08 & 0.17 ± 0.18 & — & marginal & B \\
ASW2DF.44 & 11:40:48.99 & -26:30:17.1 & 4.51 & 4.14 & 0.26 ± 0.06 & 0.56 ± 0.12 & 0.90 ± 0.15 & — & cloud-like & B \\
ASW2DF.45 & 11:40:53.96 & -26:29:00.9 & 4.43 & 4.15 & 0.21 ± 0.05 & 0.27 ± 0.06 & 0.35 ± 0.13 & — & — & B \\
ASW2DF.46 & 11:40:50.71 & -26:28:31.4 & 4.35 & 4.08 & 0.25 ± 0.06 & 0.26 ± 0.08 & 0.32 ± 0.20 & — & — & B \\
ASW2DF.47 & 11:40:51.11 & -26:29:05.1 & 4.24 & 4.05 & 0.25 ± 0.06 & 0.48 ± 0.10 & 0.61 ± 0.15 & CO$_{\rm ext}$ & possible & B \\
\bottomrule
\end{tabularx}
\tablefoot{(1) ALMA sources with IDs sorted by the peak S/N; (2) Right Ascension (J2000); (3) Declination (J2000); (4) S/N at the peak pixel; (5) S/N at the fifth brightest pixel of a source, which can be detected by the ``fixed'' method at 4$\sigma$; (6) Peak flux at 1.2\,mm; (7) Gaussian fitting flux; (8) 2$\times$synthesized beam aperture flux; 
(9) Counterparts of ALMA sources matched from different galaxy populations including SMGs \citep{Dannerbauer2014}, HAEs \citep{Koyama2013, Shimakawa2018b}, and CO emitters \citep{Jin2021, Chen2024}; CO emitters identified as extended gas reservoirs by \citet{Chen2024} are marked with a subscript `ext'; (10) The classifications of the sources identified by visual inspection, including `robust', `possible', `could-like' and `edge'; (11) source category.}
\end{table*}

\begin{table*}[ht]
\centering
\caption{Supplementary source catalog.} 
\label{table: supp catalog}
\small 
\begin{tabularx}{\textwidth}{ccccccccccc}
\toprule
ID & R.A. & Dec. & S/N$_{\rm p}$ & S/N$_{\rm d}$ & $S_{\rm peak}$ & $S_{\rm fit}$ & $S_{\rm aper}$ & Counterpart & eyes ID  & Cat \\
& (J2000) & (J2000) & & & (mJy) & (mJy) & (mJy) & & & \\
(1) & (2) & (3) & (4) & (5) & (6) & (7) & (8) & (9) & (10) & (11) \\
\midrule
ASW2DF.48 & 11:40:34.69 & -26:28:49.4 & 4.86 & 3.96 & 0.22 ± 0.04 & 0.15 ± 0.04 & 0.34 ± 0.10 & — & possible & C \\
ASW2DF.49 & 11:40:44.57 & -26:30:05.7 & 4.84 & 3.90 & 0.20 ± 0.04 & 0.12 ± 0.04 & 0.10 ± 0.11 & — & — & C \\
ASW2DF.50 & 11:40:54.51 & -26:27:47.6 & 4.52 & 3.80 & 0.35 ± 0.08 & 0.24 ± 0.07 & 0.30 ± 0.19 & — & marginal & C \\
ASW2DF.51 & 11:40:57.76 & -26:29:24.3 & 4.47 & 3.96 & 0.21 ± 0.05 & 0.16 ± 0.05 & 0.15 ± 0.13 & — & — & C \\
ASW2DF.52 & 11:40:45.22 & -26:29:20.4 & 4.46 & 3.79 & 0.18 ± 0.04 & 0.14 ± 0.04 & 0.07 ± 0.12 & — & — & C \\
ASW2DF.53 & 11:40:38.03 & -26:30:04.0 & 4.45 & 3.93 & 0.23 ± 0.05 & 0.19 ± 0.07 & 0.07 ± 0.14 & — & possible & C \\
ASW2DF.54 & 11:40:53.70 & -26:30:41.2 & 4.45 & 3.93 & 0.22 ± 0.05 & 0.20 ± 0.05 & 0.21 ± 0.12 & — & — & C \\
ASW2DF.55 & 11:40:35.01 & -26:29:07.5 & 4.45 & 3.88 & 0.20 ± 0.04 & 0.14 ± 0.05 & 0.07 ± 0.11 & — & — & C \\
ASW2DF.56 & 11:40:42.25 & -26:30:48.7 & 4.44 & 3.60 & 0.28 ± 0.06 & 0.17 ± 0.05 & 0.12 ± 0.17 & — & possible & C \\
ASW2DF.57 & 11:40:58.45 & -26:30:42.5 & 4.42 & 3.94 & 0.22 ± 0.05 & 0.22 ± 0.06 & 0.24 ± 0.12 & — & — & C \\
ASW2DF.58 & 11:40:59.10 & -26:29:07.6 & 4.40 & 3.86 & 0.20 ± 0.05 & 0.18 ± 0.04 & 0.25 ± 0.12 & — & — & C \\
ASW2DF.59 & 11:40:58.98 & -26:29:23.8 & 4.39 & 3.83 & 0.20 ± 0.05 & 0.19 ± 0.06 & 0.09 ± 0.10 & — & — & C \\
ASW2DF.60 & 11:40:36.22 & -26:27:45.9 & 4.29 & 3.97 & 0.37 ± 0.09 & 0.40 ± 0.10 & 0.28 ± 0.22 & — & marginal & C \\
ASW2DF.61 & 11:40:35.80 & -26:30:30.1 & 4.28 & 3.94 & 0.19 ± 0.05 & 0.18 ± 0.05 & 0.07 ± 0.11 & — & possible & C \\
ASW2DF.62 & 11:40:41.80 & -26:30:47.0 & 4.28 & 3.71 & 0.25 ± 0.06 & 0.32 ± 0.08 & 0.39 ± 0.17 & — & cloud-like & C \\
ASW2DF.63 & 11:40:36.70 & -26:30:37.6 & 4.21 & 3.84 & 0.19 ± 0.05 & 0.13 ± 0.04 & 0.16 ± 0.13 & — & — & C \\
ASW2DF.64 & 11:40:46.87 & -26:29:36.9 & 4.09 & 3.77 & 0.38 ± 0.09 & 0.50 ± 0.12 & 0.58 ± 0.25 & — & marginal & C \\
ASW2DF.65 & 11:40:38.38 & -26:28:53.8 & 4.02 & 3.75 & 0.25 ± 0.06 & 0.18 ± 0.05 & 0.29 ± 0.17 & — & marginal & C \\
ASW2DF.66 & 11:40:46.91 & -26:30:52.1 & 4.00 & 3.60 & 0.24 ± 0.06 & 0.22 ± 0.07 & 0.17 ± 0.19 & — & possible & C \\
ASW2DF.67 & 11:40:41.69 & -26:29:40.0 & 3.93 & 3.60 & 0.19 ± 0.05 & 0.28 ± 0.07 & -0.00 ± 0.10 & — & possible & C \\
ASW2DF.68 & 11:40:40.05 & -26:29:03.7 & 3.83 & 3.28 & 0.19 ± 0.05 & 0.45 ± 0.12 & 0.58 ± 0.10 & — & cloud-like & C \\
\bottomrule
\end{tabularx}
\tablefoot{The columns are the same as those in Table \ref{table: main catalog}.}
\end{table*}

Furthermore, an independent visual inspection was conducted on the ALMA maps to verify the source catalogs extracted by the different codes and identify faint, extended sources below the detection limit \citep{Gonzalez2020}. We visually identified 53 sources on the six ALMA maps, bypassing the use of any source extraction code. We further classified the 53 sources into four categories including ``robust'' --- bright and prominent sources in the entire map; ``possible'' --- obviously yet less bright sources with a high possibility of being genuine; ``cloud-like'' --- sources exhibiting extended features; ``edge'' --- sources resembling those in the ``possible'' category but located at the mosaic edges in each field. Among the 53 sources identified by our visual inspection, we find that 41 are included in the main catalog, leaving 12 eye-selected sources undetected by the ``fixed'' method. These 12 sources are also marked as tentative.  Additionally, there is a small overlap: 13 tentative sources identified by the ``individual'' method include some of the 12 visually inspected tentative sources. By merging these two catalogs we obtain a supplementary catalog including 21 tentative sources in total. Notably, there are 16 sources identified as ``robust'' by visual inspection, all of which are included in the main catalog.  Additionally, we note that all sources above 5$\sigma$ in the main catalog are recovered by visual inspection, suggesting the reliability of the visual inspection.

\begin{figure*}[h!]
\sidecaption
\includegraphics[width=5.9cm]{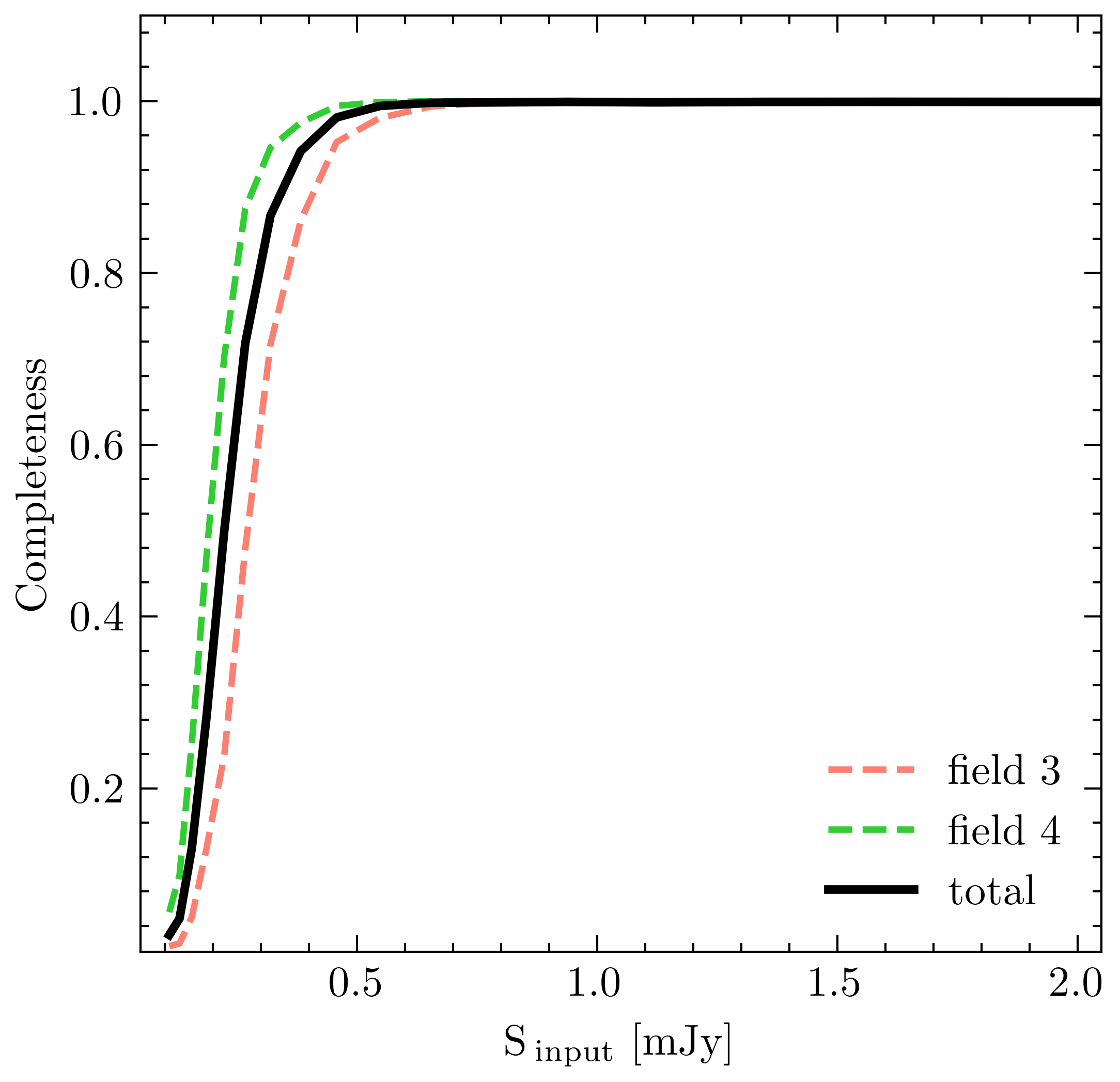}
\includegraphics[width=6cm]{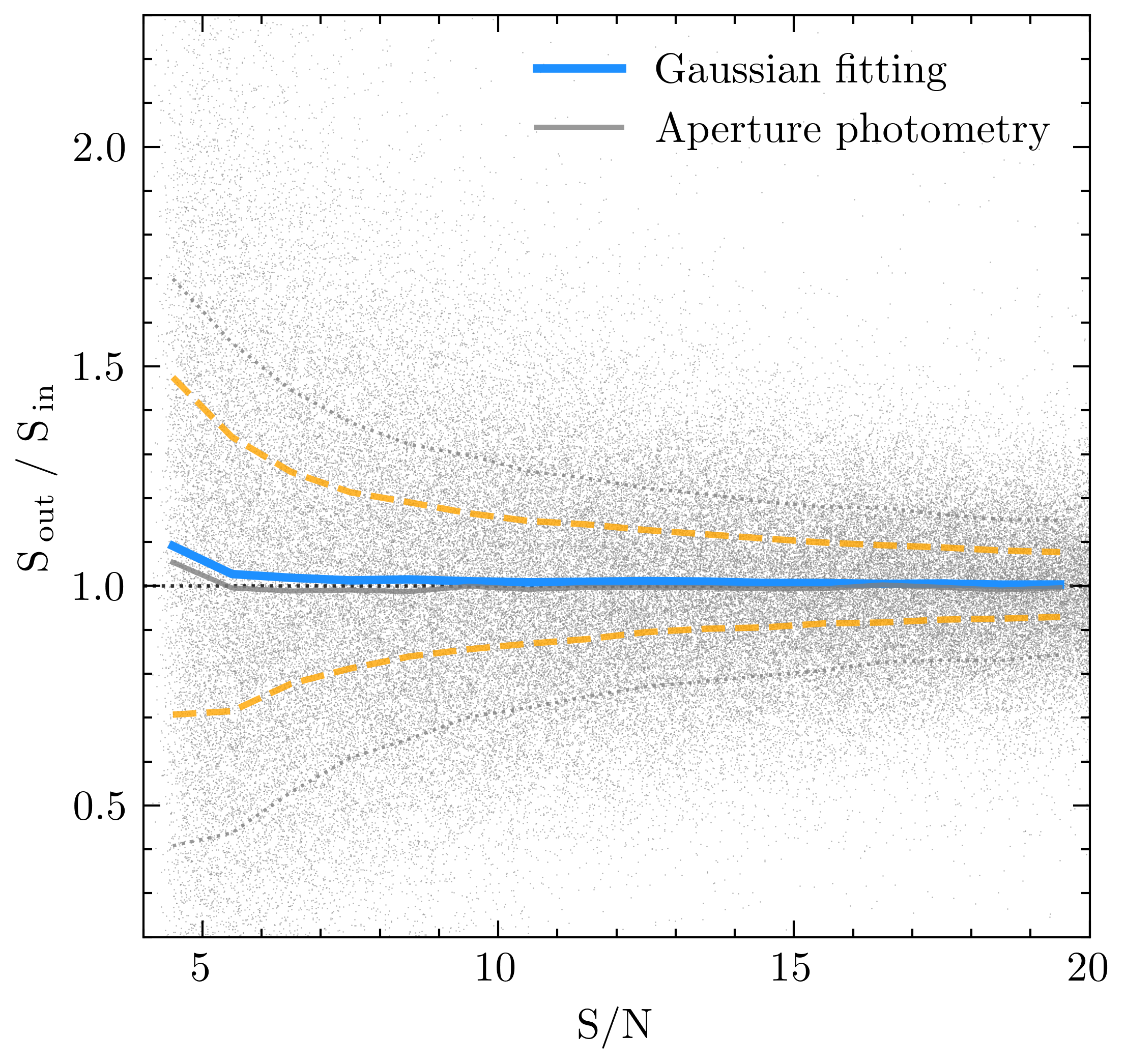}
\caption{{\bf Left:} Completeness as a function of input fluxes obtained from the MC simulation for all the six fields (black). The green and salmon lines show the completeness for Field 3 and 4, which are slightly different from the combined one and do not affect the results of the analysis. {\bf Right:} Flux boosting as a function of peak S/N. The solid lines represent the median value of the boosting factor from the Gaussian fitting (blue) and aperture photometry (gray). The yellow dashed lines and gray dotted lines indicate uncertainties from the Gaussian fitting and aperture photometry, respectively. \\ \\}
     \label{fig: comp_boost}
\end{figure*}

To summarize, we classify the 16 bright sources identified by visual inspection as ``best'' in category A, and the remaining 31 sources of the main catalog are included in category B. We note that all 47 ALMA sources in the main catalog were securely detected by the codes through the ``fixed'' method and will be used for the number counts calculations. The supplementary catalog comprises category C, encompassing 21 tentative sources identified through visual inspection and/or detected by codes using the ``individual'' method. The tentative sources in the supplementary catalog must be taken with caution and will not be used in the following number counts analysis. Figure \ref{fig: detections} shows the detections from category A, B, and C overlaid on the 1.2\,mm continuum maps. 
In addition, we present our main catalog that includes 47 ALMA sources and a supplementary catalog comprising 21 tentative sources in Tables \ref{table: main catalog} and \ref{table: supp catalog} respectively, sorted by their peak signal-to-noise ratios. 

In order to examine the probability of the spurious sources in the science catalog, we calculate the purity of our images. The purity is defined as following:
\begin{equation}
    p = \frac{N_{\rm pos}-N_{\rm neg}}{N_{\rm pos}}\,,
\label{eq: purity}
\end{equation}
where $p$ is the purity in each S/N bin, $N_{\rm pos}$ and $N_{\rm neg}$ are the positive and negative detections in a specific S/N bin. We obtain the negative detections by inverting the science map (multiplying the map by $-1$) and applying the same source extraction process used to detect positive peaks in the original map. Figure \ref{fig: purity} demonstrates the number of positive and negative detections, as well as the purity within the S/N range of $3.5-5.5$ from the ``fixed'' method. The purity reaches $\sim0.9$ at a threshold of S/N=4, indicating that around five sources are fake in our main catalog in a statistic sense. Sources above 5$\sigma$ are all considered to be real, which is in line with the results of \citet{Chen2023a}. The purity estimations of the ``individual'' method with three codes are also plotted in Fig. \ref{fig: purity} for a comparison. We find that detections above 4$\sigma$ significance from the ``fixed'' method have the highest purity compared to any of the three codes using the ``individual'' method.  This further confirms the reliability of the ``fixed'' method for estimating the rms noise. The adopted purity of 0.9 suggests that sources with peak S/N between 4.0 and 5.0 in our main catalog are a mixture of real and spurious detections, which might influence the subsequent analysis. To address this, we do not only conduct the analysis based on the main catalog but also perform an additional analysis using only \textgreater5$\sigma$ sources in the following sections. The results from this subset are then compared with those from the entire main catalog to assess the robustness of our analysis.

\subsection{Flux measurement}
\label{sec: flux}
We use several methods to measure the fluxes of our ALMA sources. The typical size of ALMA detections at $\sim$1\,mm are rather compact with a median size of 0\farcs1$-$0\farcs3 \citep{Gonzalez2017, Umehata2018, Gomez2022}. As the resolution of our images is $\sim$0\farcs5 or even larger, most of the detections are not spatially resolved in principle. The peak flux can serve as a good measurement of the total flux of a source in this case, since all emission from the source is convolved in the synthesized beam, and integrated into the brightest pixel. We firstly extract the peak fluxes of the ALMA sources from the images after the primary beam correction. 

We also use the elliptical Gaussian model to fit our ALMA sources and estimate their integrated fluxes. Following \cite{Chen2023a}, we fit the sources using the Levenberg-Marquardt algorithm in the {\sc Python} package  {\sc astropy.modeling}. A cutout image with a size of 5$\times$major axis is extracted, centered on each source, and used for the fitting process. The amplitude, axial ratio, and positional angle are free to change for obtaining the best-fitting values. The integrated flux of the each source are calculated by summing up the flux density within the fitted Gaussian aperture before the correction of the synthesized beam. The uncertainty in the integrated flux for each source is estimated through the standard deviation obtained from 1000 iterations of the flux measurements using the same best-fitting Gaussian aperture at random positions.

Additionally, we measure the source fluxes using the aperture photometry, which is performed by the {\sc Python} package {\sc photutils}. The aperture size and shape are not identical in the literature. For example, the circular aperture with a diameter of 1.6$^{\prime\prime}$ and 2.0$^{\prime\prime}$ were used by \citet{Gomez2022} and \citet{Fujimoto2023} respectively. While \citet{Chen2023a} reported the scaled elliptical aperture (i.e. $2\times$~synthesized beam), which results in reliable flux measurement. Since the images in our six fields have different resolutions, we adopted the elliptical aperture of $2\times$~synthesized beam to measure the aperture fluxes for the sources in each field. We list the source fluxes measured from peak pixel, Gaussian fitting and aperture photometry, as well as the classification (eyes ID) from visual inspection in Table \ref{table: main catalog} and \ref{table: supp catalog}. We note that the detectable S/N, obtained from the ratio of the flux in the fifth brightest pixel of each source and the 1$\sigma$ noise, is also listed in the source table. The detectable S/N defines the threshold at which our method can detect each source, given the parameter $minarea=5$.
 
\section{Number counts}
\label{sec: number counts}
One of the main goals of the ASW$^2$DF survey is to obtain the number counts of the DSFGs in the Spiderewb protocluster field. The number counts derived from the initial source catalog suffer from biases such as incompleteness, purity, and flux boosting. In this section, we investigate the biases and calculate the effective area before calculating the number counts.

\subsection{Completeness and flux boosting}
In general, the source detection rate increases with the intrinsic fluxes of ALMA sources, up to 100\,percent above the flux density $\sim0.5$\,mJy \citep{Gomez2022}.  Completeness is defined as the ratio between the detected sources and the real source number within a specific intrinsic flux bin. We used a Monte Carlo simulation to characterize the completeness of our science catalog. The main idea of the simulation is to build a mock catalog containing plenty of artificial sources and then assess the performance of the source extraction on the science images. 

\begin{figure}[h!]
\includegraphics[width=\columnwidth]{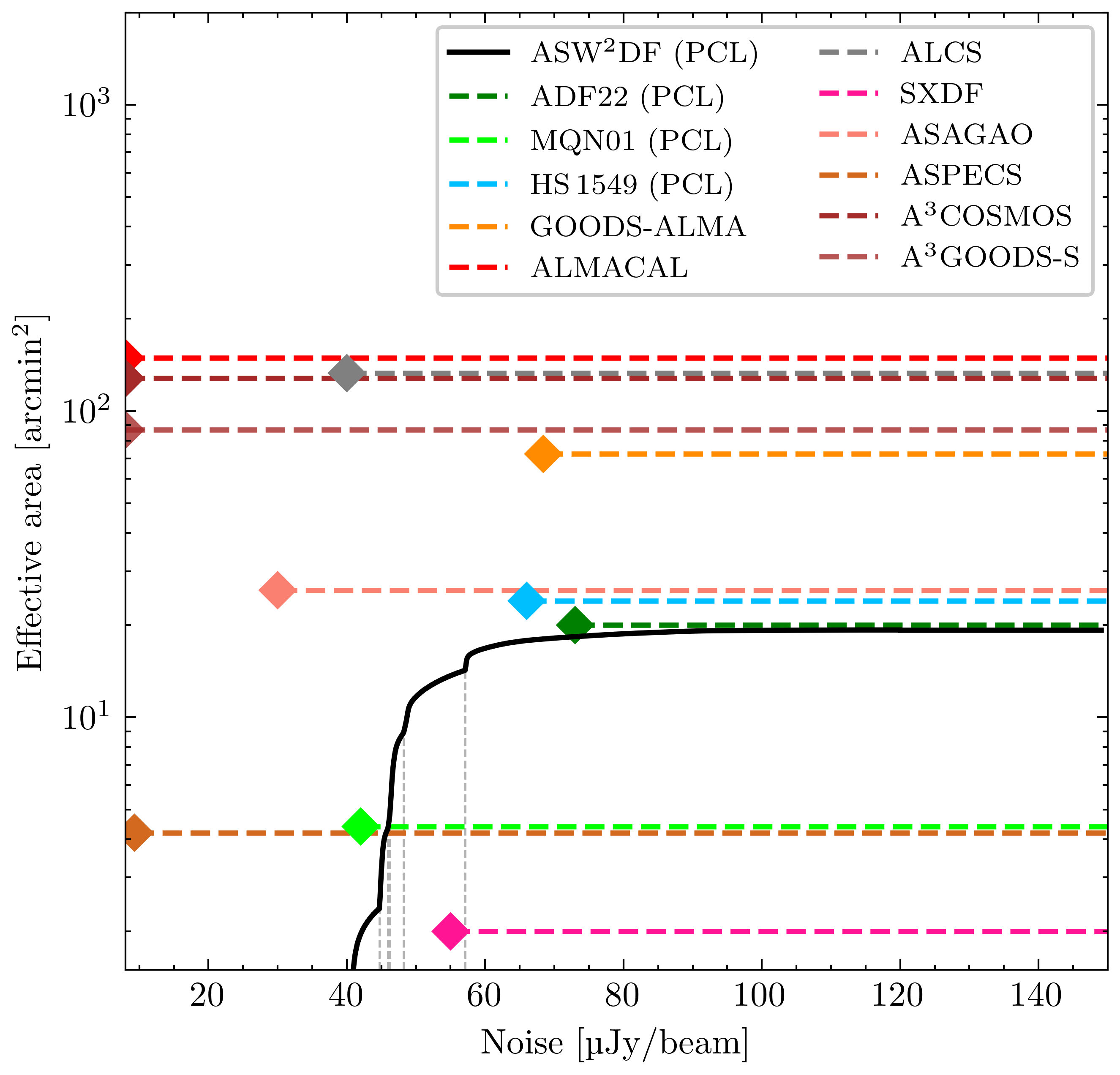}
  \caption{The effective area as a function of the local noise for our ASW$^2$DF survey. The maximum area is 19.3\,arcmin$^2$ at a maximum noise of 120\,$\mu$Jy. The discontinuities in the black curve are due to the different 1$\sigma$ sensitivities over the six fields, which are marked as vertical gray dashed lines representing Field 4, 6, 2, 1, 5, and 3 from left to right. The colored dashed lines show the survey area of other ALMA public surveys at similar wavelength, including SXDF \citep{Hatsukade2016}, ASAGAO \citep{Hatsukade2018}, ASPECS \citep{Aravena2016, Gonzalez2020}; GOODS-ALMA \citep{Gomez2022}, ALMACAL \citep{Oteo2016, Chen2023a}, A$^3$COSMOS and A$^3$GOODS-S \citep{Adscheid2024} for blank fields; ADF22 \citep{Umehata2017, Umehata2018}, MQN01 \citep{Pensabene2024} and HS\,1549 \citep{Wang2024} for protoclusters (PCL). The markers at the left side of the lines represent their typical 1$\sigma$ noise. We put the markers from ALMACAL, A$^3$COSMOS and A$^3$GOODS-S to the far left of the figure, as they do not have a representative sensitivity due to the collection of different ALMA projects.
  }
     \label{fig: area}
\end{figure}

Firstly, the 47 sources from our main science catalog (category A and B) were masked with the scaled elliptical aperture on the science images. Secondly, we divided the flux range from 0.1 to 3.1\,mJy into 30 flux bins with a step of 0.1\,mJy. In each flux bin, we generate an artificial catalog with 20 sources with assigned fluxes from the normal random distribution. The sources are positioned at random positions on the images where the primary beam response is greater than 0.5. We checked the source positions and removed the ones which are too close (i.e., less than 1\,arcsec) to avoid additional bias from source blending. Thirdly, the point-like artificial sources were convolved with the synthesized beam and then injected into the image with primary beam correction, as this image reflects the intrinsic flux of the sources. We introduced point-like sources instead of sources with specific sizes \citep[e.g., $0\farcs1-0\farcs4$;][]{Gomez2022}, given the inherent complexity in estimating the intrinsic sizes of our ALMA sources, which may vary from those typically observed in general fields. Employing this conservative approach with point-like sources leads to a slight increase in completeness \citep{Gomez2022, Chen2023a} and ensures the robustness of subsequent overdensity analysis. We multiplied the newly created image by the primary beam response to obtain the PB uncorrected image. The source extraction and flux measurement were performed on this artificial image in the same manner as we did in Sec. \ref{sec: source extraction}. The artificial sources were considered recovered when they were extracted less than 1\,arcsec from their original injected positions. The procedure was repeated for 1000 iterations in each flux bin and each field. Finally, we obtained the mock catalog with tens of thousands of recovered artificial sources, which will be used for the completeness assessment.

Figure \ref{fig: comp_boost} demonstrates the completeness result from our mock catalog. The completeness over all six fields increases rapidly within the input flux range of 0.1$-$0.5\,mJy, reaching 80\% and 100\% completeness at around 0.3\,mJy and 0.6\,mJy, respectively. As the 1$\sigma$ sensitivities vary between different fields (see Table \ref{tab: obs}), we also show the completeness estimation for Field 3 and 4 as they have the highest and lowest typical noise. It is clear that the completeness from these two fields follows a trend similar to the overall one, but is slightly different. We note that for the rest of the four fields, the completeness behaves the same as the one for all six fields and the discrepancies from Field 3 and 4 do not affect our results. We used the overall one for all six fields to correct for our science catalog.

The flux boosting effect can also be quantified through the MC simulation \citep{Hogg1998, Murdoch1973}. This has been discussed in previous ALMA studies \citep[e.g.,][]{Hatsukade2016, Umehata2017, Umehata2018, Hatsukade2018, Franco2018, Gomez2022, Chen2023a}. There are two independent reasons responsible for boosting fluxes \citep{Casey2014}. One is the so-called Eddington bias \citep{Eddington1913}. As the source number increases exponentially from the bright to the faint end, the sources scattering towards higher flux densities are much more than the number of sources scattered to lower flux densities. The other reason is that, the unresolved faint sources below the detection limit could also contribute to fluxes of the bright sources within the same beam, which are more often seen in single-dish submm/mm surveys \citep{Coppin2006}.

We used the matched catalog from the MC simulation to measure the flux boosting and assess the performance of the flux measurements for our science catalog. We obtained the flux boosting factor by dividing the recovered flux ($S_{\rm out}$) by the input flux ($S_{\rm in}$) of each source in the matched catalog. Both two-dimensional Gaussian fitting and aperture photometry are used for the flux measurement on the recovered sources. We show the flux boosting estimation from these two methods as a function of the peak S/N in Figure \ref{fig: comp_boost}. We notice that both two-dimensional Gaussian fitting and aperture photometry can recover the source flux accurately at \textgreater5$\sigma$, within a 3\,percent difference than intrinsic flux. At a peak S/N lower than five, both methods overestimate the source flux, in which a slightly higher boosting factor is shown in the Gaussian fitting method. This result is consistent with previous studies on similar comparison \citep{Gomez2022, Chen2023a}.

We note from Fig. \ref{fig: comp_boost} that the aperture photometry suffers more from the scatters of the boosting factors in all S/N bins. This implies that the measured flux may deviate further from the intrinsic flux of the individual source. Furthermore, the flux is not significantly boosted (up to 10\%) in both methods even in the lowest S/N region (4$\sigma$), rather lower than the flux uncertainty which is at a level of $\sim20$\,percent. The Gaussian fitting fluxes are adopted in the following analysis, including the number counts calculation. We also tested the analysis with aperture photometry and obtained consistent results.

\subsection{Effective area}
The science images produced by the interferometric observations are heavily influenced by the primary beam response. Our contiguous mapping observations have a better homogeneous sensitivity than the single pointing ones, but the noise still changes rapidly at the edge of each field. We calculate the effective area as a function of the local noise that increases from the center to the edges of the images. The local noise is calculated by dividing the ``fixed'' noise values by the primary beam response over the whole image where PB\textgreater0.5. Thus, the local noise ranges from one to two times the ``fixed'' rms noise value in each field, as the primary beam correction decreases from 1 to 0.5.

\begin{figure*}
\begin{center}
\includegraphics[width=\columnwidth]{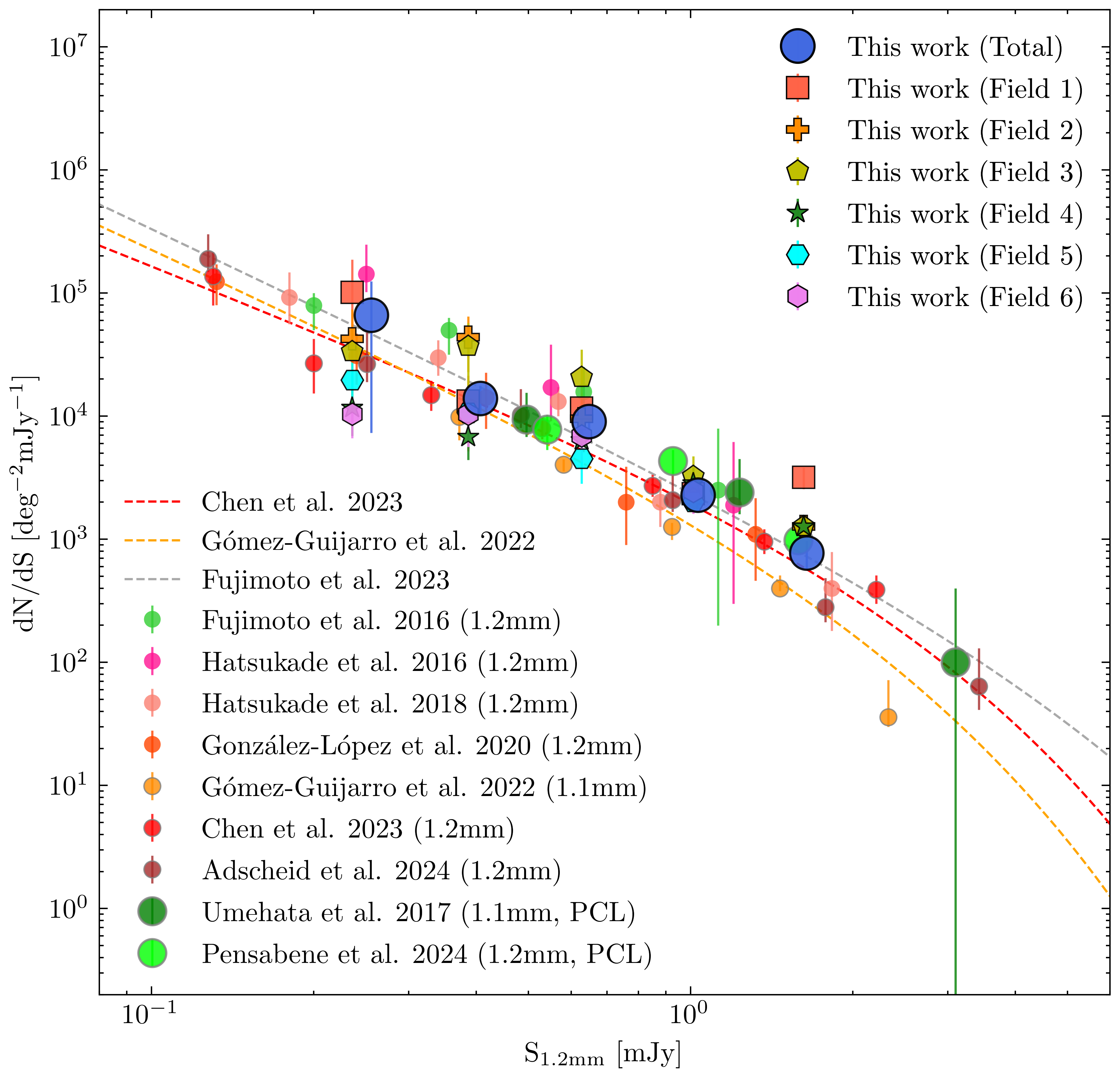}
\includegraphics[width=\columnwidth]{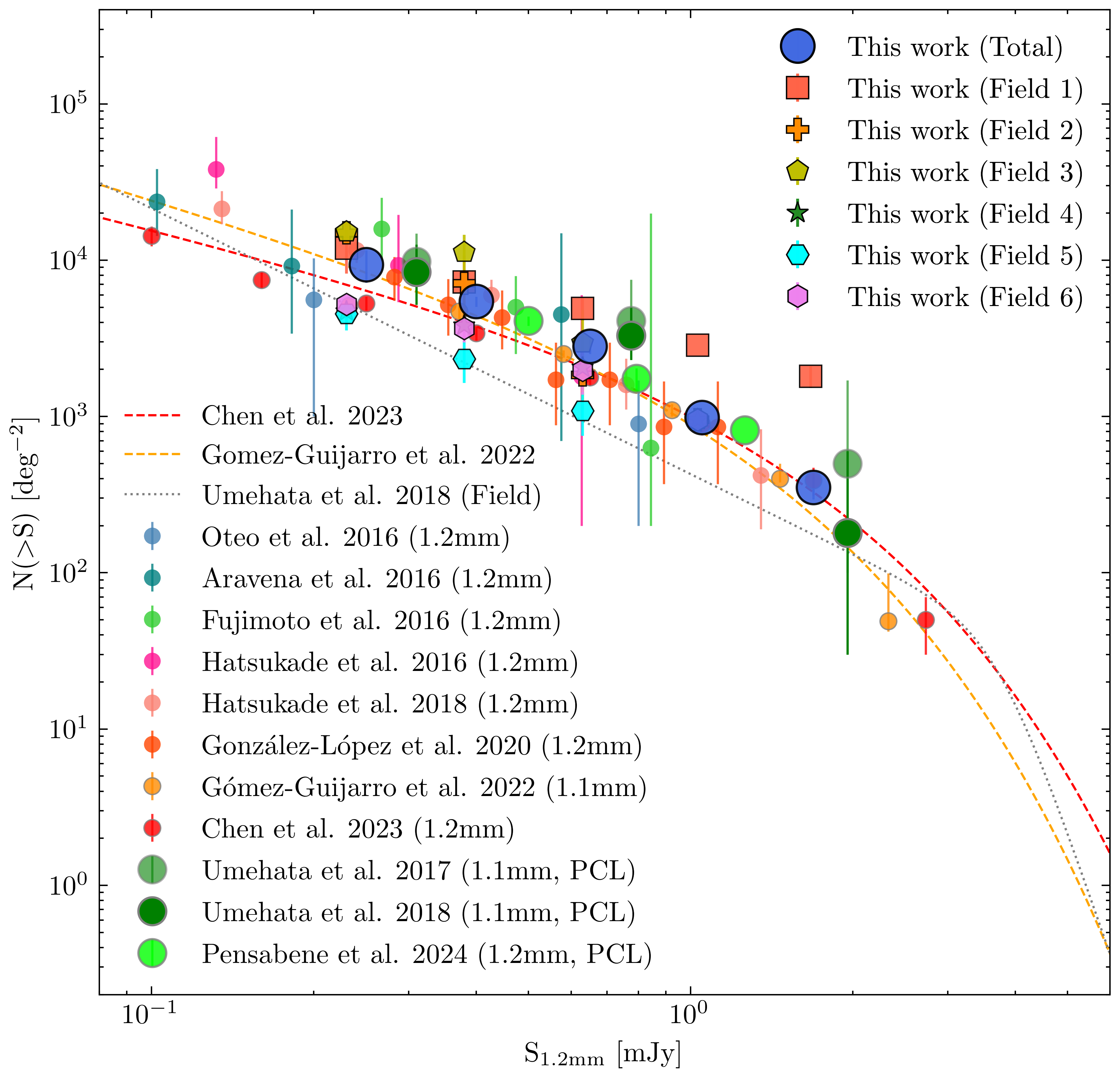}
\caption{ Differential (left) and cumulative (right) number counts for the ASW$^2$DF survey. The number counts are derived from the main science catalog with the correction of completeness and purity for each source. Big blue circles show the total number counts from the six fields, and the results of each field are shown in various markers as shown in legend. The number counts from previous surveys at similar wavelengths are also shown with colored circles, in which the larger circles represent the studies towards protocluster fields \citep[PCL:][]{Umehata2017, Umehata2018, Pensabene2024}. The gray, orange and red dashed lines show the best-fitting results in blank fields \citep{Fujimoto2023, Gomez2022, Chen2023a} with a Schechter function. The gray dotted line in the right panel represents the best-fitting of cumulative number counts in general fields used in \citet{Umehata2018}. The flux bins of the studies at 1.1\,mm are converted to that at 1.2\,mm according to the conversion factor of 1.29 (see Sec. \ref{sec: nc}).}
\label{fig: nc}
\end{center}
\end{figure*}

Figure \ref{fig: area} shows the effective area versus the local rms noise in the combination of all six fields. The total effective area increases with the local rms noise and reach the maximum of 19.3\,arcmin$^2$ at the noise of 120\,$\mu$Jy, indicating the total area of the ASW$^2$DF survey.  We note that the effective area curves are not as smooth as seen in the literature \citep[e.g.,][]{Gomez2022}, showing several discontinuities due to the different depths in the six fields. We also show the 1$\sigma$ noise for six fields with vertical dashed lines, clearly demonstrating where the discontinuities start. For a comparison, we also list the survey area from the literature targeting both fields and protoclusters \citep{Oteo2016, Aravena2016, Hatsukade2016, Hatsukade2018, Umehata2017, Umehata2018, Gonzalez2020, Gomez2022, Chen2023a, Fujimoto2023, Pensabene2024, Adscheid2024, Wang2024} in Fig. \ref{fig: area}. We choose these surveys because they were conducted at 1.1/1.2\,mm, which are very close to our observations, minimizing the uncertainty due to the flux conversion (see Sec. \ref{sec: nc}). These surveys will be used for the following discussion on number counts and galaxy overdensity. We note that our survey reaches a good balance between survey area and sensitivity compared to the archival ALMA maps. Notably, the ADF22 survey observed the protocluster SSA22 at a comparable sensitivity and survey area as ours, offering a good chance for a comparison on the ALMA-detected DSFGs in protoclusters. We calculate the effective area for each source in our catalog before constructing the number counts. For each source, we divide the peak flux by the detection threshold of four to obtain the detectable noise, the effective area of this source is derived from the detectable noise and the curve in Fig. \ref{fig: area}. The brightest sources can be detected anywhere within the survey area, thus having the maximum effective area. The faint sources can only be detected at 4$\sigma$ where the local noise is very low (i.e., the central region of each field), thus having a relatively smaller effective area.

\subsection{Number counts}
\label{sec: nc}
We derive the number counts from the main science catalog (category A and B), including 47 sources in total. Each source has a contribution to the number counts as:

\begin{equation}
\label{eq: contribution}
\xi(S) = \frac{p(S)}{A_{\rm{eff}}(S)~C(S)}\,,
\end{equation}

where $S$ is the source flux, $p(S)$ is the purity which is defined in Eq. \ref{eq: purity}. $C(S)$ is the completeness associated with the source flux. $A_{\rm{eff}}$ is the effective area for the source flux as described in the previous subsection. 
The sum of the contribution of the sources within a specific flux bin is the differential number counts:

\begin{equation}
\label{eq: diff_nc}
\frac{dN}{dS} = \frac{\sum \xi(S)}{\Delta S}\,.
\end{equation}

By summing up the contribution of sources that are brighter than the flux $S_0$, we can obtain the cumulative number counts:

\begin{equation}
\label{eq:cum_nc}
N(>S_0) = \sum \xi(S)\,.
\end{equation}

We construct the differential and cumulative number counts over the whole flux range from our main catalog. We calculate the uncertainties of the number counts by constructing 1000 rounds of the number counts. In each iteration, the flux of each source is randomly sampled based on its measured flux and the 1$\sigma$ noise. The new set of source fluxes are split into different flux bins for calculating the number counts. The median values of the 1000 rounds are adopted as the final number counts, and the 1$\sigma$ dispersion are taken as the number counts errors. We use the same approach to calculate the differential and cumulative number counts as well as their scatters, respectively. 

\begin{table*}[ht]
\centering
\caption{Summary of recent ALMA number counts surveys.}
\label{tab: surveys}
\small
\begin{tabularx}{\textwidth}{ccccccccc}
\toprule
Survey & Type & $\lambda$ & Resolution  & Depth & S/N  & N$_{\rm source}$ & Area & Ref. \\
& & (mm) & ($^{\prime\prime}$) & ($\mu$Jy) & & & (arcmin$^2$) & \\
\midrule
ASW$^2$DF & M & 1.2 & $0.33-0.94$ & $40.3-57.1$ & 4.0 & 47 & 19.3 & This work \\
A$^3$COSMOS & S & $0.45-3$ & $-$ & $-$ & 5.4 & 2204 & 896.3 & \citealp{Adscheid2024}$^a$ \\
A$^3$GOODS-S & S & $0.45-3$ & $-$ & $-$ & 5.4 & 491 & 149.6 & \citealp{Adscheid2024}$^b$ \\
MQN01 (PCL) & M & 1.26 & $1.23\times1.05$ & 42 & 4.0 & 11 & 4.4 & \citealp{Pensabene2024} \\
ALCS & M & 1.2 & $1.26\times0.98$ & 63.0 & 5.0 & 180 & 131.2 & \citealp{Fujimoto2023}$^c$ \\
ALMACAL IX & S &  $0.87-3$ & $-$ & $-$ & 5.0 & 132 & 149.0 & \citealp{Chen2023a}$^d$ \\
GOODS-ALMA 2.0 & M & 1.1 & $0.45\times0.42$ & 68.4 & 5.0 (3.5) & 44 (88) & 72.4 & \citealp{Gomez2022} \\
ASPECS-LP & M & 1.2 & $1.53\times1.08$ & 9.3 & 4.2 & 4.3 & 35 & \citealp{Gonzalez2020}$^e$ \\
ASAGAO & M & 1.2 & $0.59\times0.53$ & 30 & 5.0 (4.0) & 25 (45) & 26 & \citealp{Hatsukade2018} \\
ADF22 (PCL) & M & 1.1 & $0.53\times0.52$ & 73 & 5.0 (4.5) & 35 (58) & 20 & \citealp{Umehata2018}$^g$ \\
ADF22 (PCL) & M & 1.1 & $1.16\times1.02$ & 60 & 5.0 & 18 & 6 & \citealp{Umehata2017}$^h$ \\
ALMACAL I & S & 1.2 (0.87) & \textless1 & $\sim$250 & 5.0 & 8 (11) & 16 (6) & \citealp{Oteo2016} \\
ASPECS & M & 1.2 (3.0) & $1.70\times0.90$ & 12.7 & 3.5 & 9 & 1 & \citealp{Aravena2016}$^i$ \\
ALMA Archive & S & 1.2 & $-$ & $-$ & 3.0 & 122 & 8.9 & \citealp{Fujimoto2016} \\
SXDF–ALMA & M & 1.1 & $0.53\times0.41$ & 55 & 4.0 & 23 & 2 & \citealp{Hatsukade2016} \\
\midrule
Ex-MORA & M & 2.0 & $1.68\times1.44$ & 51 & 5.0 & 37 & 577 & \citealp[]{Long2024} \\
MORA & M & 2.0 & $1.80\times1.40$ & 90 & 5.0 & 13 & 184 & \citealp{Zavala2021} \\
AS2COSMOS & S & 0.87 & $0.80\times0.79$ & 190 & 4.8 & 260 & 5760$^f$ & \citealp{Simpson2020} \\
ALMACAL VII & S & 0.65 & $0.34-0.98$ & $47-1022$ & 4.5 & 21 & 5.5 & \citealp{Klitsch2020} \\
AS2UDS & S & 0.87 & $0.15-0.30$ & 250 & 4.3 & 695 & 50 & \citealp{Stach2018} \\
ALMA Archive & S & 3.0 & \textgreater1.0 & \textless200 & 5.0 & 16 & 200 & \citealp{Zavala2018} \\
GOODS-ALMA & M & 1.1 & $0.61\times0.59$ & 182 & 4.8 & 20 & 69 & \citealp{Franco2018} \\
ALMA Frontier Fields & S & 1.1 & $0.50-1.50$ & 55-71 & 4.5 & 19 & 14 & \citealp{Munoz2018} \\
ALMA HUDF & M & 1.3 & $0.71\times0.67$ & 34 & 3.5 & 16 & 4.5 & \citealp{Dunlop2017} \\
ALMA UKIDSS UDS & S & 0.87 & $0.80\times0.65$ & 260 & 4.0 & 52 & 2808 & \citealp{Simpson2015}$^j$ \\ 
ALMA Archive & S & 1.1 (1.3) & $-$ & $7.8-52.1$ & 3.5 & 50 & 3.4 (2.4) & \citealp{Carniani2015} \\
ALMA Archive & S & 1.2 & $0.6-1.5$ & $17-88$ & 4.0 & 11 & 2.9 & \citealp{Ono2014} \\
ALMA Archive & S & 1.3 & $0.6-1.3$ & $40-100$ & 3.8 & 15 & 2.7 & \citealp{Hatsukade2013} \\
ALESS & S & 0.87 & $1.60\times1.15$ & 400 & 3.5 & 32 & $-$ & \citealp{Hodge2013}$^k$ \\
ALESS & S & 0.87 & $1.80\times1.20$ & \textless600 & 3.5 & 99 & 880 & \citealp{Karim2013} \\
\bottomrule
\end{tabularx}
\tablefoot{The top panel shows the references we used for the discussion of number counts and galaxy overdensity. Surveys targeting protoclusters are marked with ``PCL'' in brackets. Depending on the observing mode, contiguous mapping surveys are marked as ``M'' and  single pointings or the collections of archival surveys are marked as ``S''. (a\&b): The S/N threshold and source number are adopted from the blind search category, the survey area is the combined area from band 3 to band 9; (c): The resolution and depth are adopted from the average of 33 fields; (d): The resolution and depth are various between observations, the source number and survey area is adopted from band 6 at 1.2\,mm; (e): The resolution and depth are from the natural weighted data; (f): The area is from the S2COSMOS survey \citep{Simpson2019} and used for constructing their number counts; (g): The resolution and depth are the mean values from the 0\farcs60 tapered images; (h): The resolution is from the FULL/LORES data, which was primarily used in their analysis; (i): The resolution, depth and source number are adopted from 1.2\,mm data; (j): The resolution and depth are adopted from the low resolution ``detection'' maps, the survey area is from the parent single-dish survey; (k): The resolution and depth are adopted from the median value from the ALESS pointings.
}
\end{table*}

Figure \ref{fig: nc} shows the number counts for each field, as well as all six fields together. The flux bins for each field are the same with that from the total fields but shifted for 0.02\,mJy to avoid the overlap. We note that in the main catalog, there are only six sources brighter than 1\,mJy and one brighter than 2\,mJy. This aligns with the source density in GOODS-ALMA \citep{Gomez2022}, where 18 (7) sources brighter than 1 (2) mJy are found in a survey area four times larger than that of our ALMA maps. However, there are 16 sources brighter than 1\,mJy found in the ADF22 survey after accounting for the flux conversion \citep{Umehata2018}. The number of bright sources is around three times higher than compared to the 6 bright sources (\textgreater1\,mJy) in the Spiderweb field, within a similar survey area of $\sim20$\,arcmin$^2$. We argue that this is because the DSFGs are located near the gas rich filaments in SSA22, which has a relative earlier evolutionary stage at higher redshift \citep{Umehata2019}. Huge amount of gas is accreted into these galaxies and tremendous star formation is triggered subsequently.

\begin{figure*}[h!]
\centering
\includegraphics[width=\textwidth]{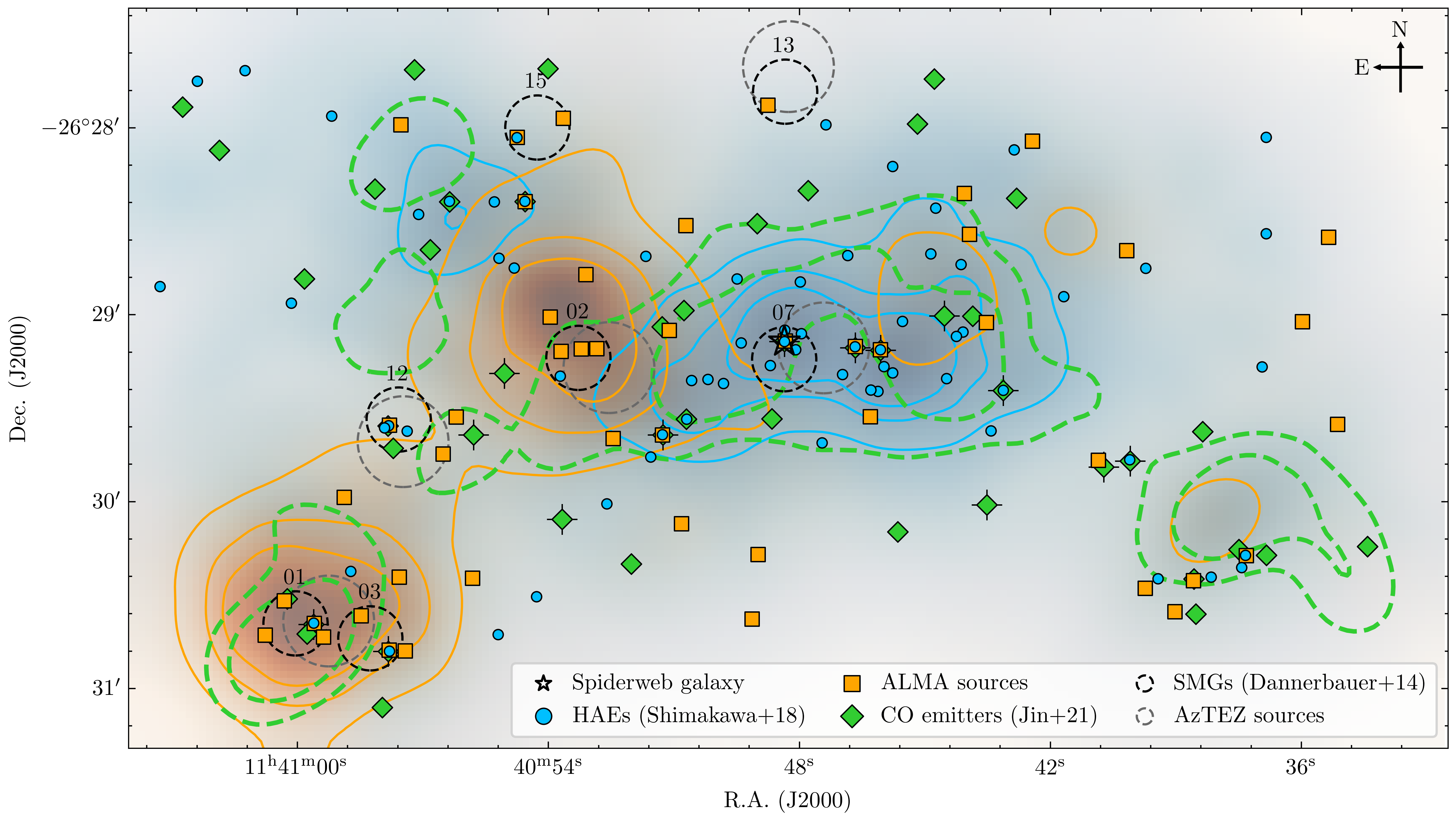}
\caption{Spatial distribution of the ALMA sources detected in the ASW$^2$DF survey. The blue circles mark the HAEs overlaid on the corresponding blue density map, which shows the excess of the surface number densities based on the 5th neighbour analysis \citep{Shimakawa2018b}. Our 47 ALMA sources  from  the main catalog are shown in orange squares, overlaid on the orange density map with same method as that of HAEs. 
We also show 46 CO emitters detected in the COALAS survey \citep{Jin2021} as green diamonds, green dashed contours represent the corresponding density maps similar to those of ALMA sources and HAEs. Furthermore, 14 CO emitters identified as robust candidates for extended molecular gas reservoirs \citep{Chen2024} are highlighted with black crosses. Seven LABOCA sources \citep{Dannerbauer2014} and five AzTEC sources \citep{Zeballos2018} within the ALMA field of view are shown as black dashed circles and gray dashed circles, which are sized according to the beam sizes of their respective single-dish telescopes. }
\label{fig: density map}
\end{figure*}

Our ASW$^2$DF survey targets the prominent Spiderweb protocluster at $z=2.16$, therefore, it is important to compare our results to the number counts in general fields and other protoclusters at similar wavelengths from the literature. In Fig. \ref{fig: nc}, we plot the results of ALMA observations at 1.1/1.2\,mm from the literature \citep{Oteo2016, Aravena2016, Fujimoto2016, Hatsukade2016, Hatsukade2018, Umehata2017, Umehata2018, Gonzalez2020, Gomez2022, Chen2023a, Pensabene2024, Adscheid2024} as we mentioned in previous section. The central wavelength of our observations is 1.2\,mm, and for the number counts at 1.1\,mm wavelength we re-scale the flux density to make the number counts comparable with our results. We adopt an average spectral energy distribution from a sample of ALESS SMGs \citep{da_Cunha2015}, and assume a redshift of 2.2 which is close to that of the Spiderweb protocluster. We then obtain the flux conversion factor of $S_{\rm 1.1\,mm}/S_{\rm 1.2\,mm}$ to be 1.29, identical to that in \citet{Umehata2018} by using a modified black body model. We note that even at the same wavelength, the number counts from different surveys show a scatter due to the variation of the survey designs, observing depth and data analysis.

By comparing our results with the field number counts in the literature as shown in Fig. \ref{fig: nc}, we find that both differential and cumulative number counts in the entire Spiderweb field are higher than those in general fields, especially at the faint end. Qualitatively, it is clearly seen that the cumulative number counts in Field 1 to 3 are overdense than that of general fields, while Field 4 to 6 show comparable or less dense signatures. We notice that the cumulative number counts in Field 1 are extraordinarily higher than in the other fields at the bright end, as three out of six bright sources (\textgreater1\,mJy) are in Field 1.
Among the previous ALMA contiguous mapping surveys --- only a few targeted on protocluster regions \citep{Umehata2015, Umehata2017, Umehata2018, Pensabene2024} ---, we also plot the number counts in the field of the protocluster SSA22 at $z=3.1$, which shows an excess by a factor of three to five at 1.1\,mm  \citep{Umehata2017, Umehata2018}. For the protocluster around the quasar MQN01, we adopt their source catalog and construct the number counts after accounting for the purity \citep{Pensabene2024}, as shown in Fig. \ref{fig: nc}. We can see that our results are comparable with the number counts in SSA22 \citep{Umehata2018} and MQN01 \citep{Pensabene2024}. We will calculate and discuss the overdensity of ALMA sources quantitatively in the next section. 

\section{Discussion}
\label{sec: discussion}
\subsection{Overdensity of DSFGs detected by ALMA}
\label{sec: overdensity}
Based on the 47 ALMA sources in our main science catalog, we examine if there is an excess of DSFGs in the Spiderweb protocluster. In Table \ref{tab: surveys}, we summarize the surveys conducted in the last decade, constructing ALMA number counts mostly in bands 6 and 7. We collected the source catalogs from field surveys \citep{Oteo2016, Aravena2016, Fujimoto2016, Hatsukade2016, Hatsukade2018, Umehata2017, Umehata2018, Gonzalez2020, Gomez2022, Chen2023a, Pensabene2024, Adscheid2024} as listed in the upper panel of Table \ref{tab: surveys}, which are also plotted in Fig. \ref{fig: nc}. For the sake of fairness, we only count the sources brighter than 0.23\,mJy at 1.2\,mm, to ensure a reliable detection of each source at \textgreater4$\sigma$ significance, given the sensitivity of $\sim$50\,$\mu$Jy in our observations as well as most archival surveys at 1.2\,mm. Similarly, a flux lower limit of 0.3\,mJy is adopted for reliable detections (\textgreater$4\sigma$) in the surveys at 1.1\,mm, which mostly have $\sim$70\,$\mu$Jy sensitivity \citep{Umehata2018, Gomez2022}.  We calculate the number density of the ALMA sources in our main catalog and public surveys by simply dividing the source numbers by the survey area respectively. We obtain a number density of $\rho\sim2.1\pm0.3\,\rm arcmin^{-2}$ in the field of the Spiderweb protocluster, surpassing the average number density of $\rho\sim1.0\pm0.4\,\rm arcmin^{-2}$ in general fields. The number density of ALMA sources detected in the SSA22 field is $\rho\sim2.8\pm0.4\,\rm arcmin^{-2}$, slighly higher than that in the Spiderweb field. The overdensity of three times than fields is consistent with that reported in \citet{Umehata2017, Umehata2018}.  In the protocluster around the quasar MQN01 \citep{Pensabene2024}, the surface density is estimated to be $\rho\sim2.0\pm0.4,\rm arcmin^{-2}$, revealing an  overdensity of twice and comparable to that in the Spiderweb protocluster. The similarity of number counts between the MQN01 and Spiderweb fields can also be seen in Fig. \ref{fig: nc}.

\begin{table*}[h!]
\centering
\caption{ALMA 1.2\,mm number counts and overdensity factors in the Spiderweb protocluster field.}
\label{tab: nc}
\small
\begin{tabular}{cccccccc}
\toprule
 & Field 1 & Field 2 & Field 3 & Field 4 & Field 5 & Field 6 & Total \\
\midrule
$S$\,[mJy] & \multicolumn{7}{c}{dN/d$S$\,[10$^3$ mJy$^{-1}$deg$^{-2}$]} \\
\rule{0pt}{2ex}
$0.20-0.31$ & 
$108.27\pm88.26$ & $41.34\pm30.67$ &
$32.81\pm18.10$ & $11.33\pm4.31$ &
$20.11\pm8.39$ & $10.52\pm3.79$ &
$66.73\pm66.73$ \\
$0.31-0.50$ & 
$13.24\pm7.22$ & $43.96\pm20.69$ &
$37.14\pm20.03$ & $6.72\pm2.36$ &
$10.58\pm4.18$ & $10.20\pm4.72$ &
$14.00\pm14.00$ \\
$0.50-0.80$ & 
$11.58\pm2.36$ & $9.76\pm6.14$ &
$21.90\pm14.25$ & $6.43\pm2.04$ &
$4.61\pm1.74$ & $6.79\pm2.69$ &
$9.10\pm9.10$ \\
$0.80-1.27$ & 
$2.31\pm0.67$ & $2.08\pm0.36$ &
$3.20\pm1.43$ & $2.86\pm1.00$ &
$2.02\pm0.00$ & $2.53\pm0.89$ &
$2.33\pm2.33$ \\
$1.27-2.02$ & 
$3.19\pm0.64$ & $1.27\pm0.00$ &
$1.27\pm0.00$ & $1.27\pm0.00$ &
$-$ & $-$ &
$0.77\pm0.77$ \\
\midrule
$S$\,[mJy] & \multicolumn{7}{c}{N(\textgreater$S$)\,[10$^3$ deg$^{-2}$]} \\
\rule{0pt}{2ex}
0.25 & 
$11.93\pm3.73$ & $14.95\pm2.42$ &
$15.20\pm1.66$ & $4.69\pm0.49$ &
$4.54\pm0.99$ & $5.21\pm0.50$ &
$9.38\pm2.21$ \\
0.40 & 
$7.25\pm0.72$ & $7.13\pm3.60$ &
$11.23\pm3.29$ & $3.82\pm0.20$ &
$2.33\pm0.70$ & $3.66\pm0.69$ &
$5.44\pm0.40$ \\
0.65 & 
$4.91\pm0.81$ & $1.85\pm0.68$ &
$2.96\pm1.85$ & $2.06\pm0.39$ &
$1.09\pm0.34$ & $1.98\pm0.59$ &
$2.82\pm0.31$ \\
1.05 & 
$2.86\pm0.00$ & $0.95\pm0.00$ &
$0.96\pm0.01$ & $0.95\pm0.04$ &
$-$ & $0.95\pm0.00$ &
$0.99\pm0.09$ \\
1.69 & 
$1.81\pm0.29$ & $-$ &
$-$ & $-$ &
$-$ & $-$ &
$0.35\pm0.06$ \\
\midrule
$S$\,[mJy] & \multicolumn{7}{c}{ALMA overdensity} \\
\rule{0pt}{2ex}
0.25 & 
$2.25\pm0.74$ & $2.82\pm0.54$ &
$2.87\pm0.42$ & $0.88\pm0.13$ &
$0.86\pm0.21$ & $0.98\pm0.14$ &
$1.77\pm0.45$ \\
0.40 & 
$2.13\pm0.30$ & $2.09\pm1.08$ &
$3.29\pm1.02$ & $1.12\pm0.13$ &
$0.68\pm0.22$ & $1.07\pm0.23$ &
$1.59\pm0.20$ \\
0.65 & 
$2.76\pm0.56$ & $1.04\pm0.40$ &
$1.66\pm1.06$ & $1.16\pm0.26$ &
$0.61\pm0.20$ & $1.11\pm0.36$ &
$1.58\pm0.26$ \\
1.05 & 
$3.21\pm0.47$ & $1.07\pm0.16$ &
$1.07\pm0.16$ & $1.07\pm0.16$ &
$-$ & $1.07\pm0.16$ &
$1.11\pm0.19$ \\
1.69 & 
$4.63\pm1.20$ & $-$ &
$-$ & $-$ &
$-$ & $-$ &
$0.90\pm0.24$ \\
U18 field & $2.64\pm0.83$ & $3.31\pm0.53$ & $3.36\pm0.37$ & $1.04\pm0.11$ & $1.00\pm0.22$ & $1.15\pm0.11$ & $2.08\pm0.49$ \\ 
\midrule
& \multicolumn{7}{c}{ALMA overdensity around the SW galaxy} \\
\rule{0pt}{2ex}
Radii (kpc) & \textless150 & $150-300$ & $300-450$ & $450-600$ & $600-750$ & $750-900$ & $900-1050$ \\
Factor & 3.5 & 2.4 & 2.8 & 4.0 & 1.6 & 1.3 & 1.4 \\
\bottomrule
\end{tabular}
\tablefoot{The overdensity in each field at each flux bin is calculated based on the field results from \citet{Chen2023a} and \citet{Gomez2022}. The overdensity results with a name ``U18 field'' are calculated based on the field results from \citet{Umehata2018}.}
\end{table*}

We use the derived number counts to compare with the results from general fields and quantify the overdensity of DSFGs in the Spiderweb protocluster. We adopt the recent results from the ALMACAL \citep{Chen2023a} and GOODS-ALMA \citep{Franco2018, Gomez2022} surveys because they have the largest survey area in individual pointings and contiguous mosaic modes respectively, minimizing the effect of the selection bias and cosmic variance. \citet{Chen2023a} constructed the number counts at exactly the same wavelength as in our observations, thus avoiding the uncertainty from the flux conversion.  We convert the number counts at 1.1\,mm in GOODS-ALMA survey to that at 1.2\,mm by using the factor of 1.29 mentioned above. The discrepancies in differential number counts between bright and faint ends will also affect the overdensity estimation. Thus we use the cumulative number counts to evaluate the overdensity for the entire population of the ALMA sources. Based on the sensitivity of our science maps and the flux range of the main catalog, we adopt the five flux bins in logarithmic space within a flux range of 0.25$-$1.69\,mJy and calculate the cumulative number counts. We only compare our results with the data points in the literature instead of their best-fitting curves in order to avoid fitting errors. The overdensity is characterized by the ratio between the number counts in the Spiderweb protocluster and the results from general fields.
We note that the cumulative number counts in the A3COSMOS survey \citep{Adscheid2024} are not provided and cannot be directly compared. Nevertheless, their differential number counts are in a good agreement with that in the literature, thus the overdensity calculation would be the same if their cumulative number counts were available.

The number counts and overdensity of our ALMA sources are shown in Table \ref{tab: nc}. At the lowest flux limit of 0.25\,mJy, we estimate the number counts in the Spiderweb protocluster around two times that of general fields. This is consistent with the previous estimation mentioned above by only counting the source numbers. The overdensity is prominent in the eastern part of the Spiderweb protocluster (from Field 1 to Field 3), while the number counts in the western part (from Field 4 to Field 6) are close to that of general fields. In contrast, the HAEs are mostly concentrated around the Spiderweb galaxy (i.e., the protocluster center), with filament-like structures extending to the east and west \citep{Koyama2013, Shimakawa2018b}. Notably, Field 1 shows the highest overdensity of ALMA sources at the bright end, the number density excess increases with the flux bins reaching up to 3.2 (4.6) at \textgreater1.05 (1.69)\,mJy. This trend can also be seen in the cumulative number counts as shown in Fig. \ref{fig: nc}. In addition, we measured the overdensity factor around the Spiderweb galaxy of 3.5, 2.4, 2.8, 4.0, 1.6, 1.3 and 1.4 in radii of $<150 , 150-300, 300-450, 450-600, 600-750, 750-900, 900-1050$ kpc, respectively.

\citet{Umehata2018} reported that the number counts of 1.1\,mm ALMA sources in SSA22 have an overdensity by a factor of three to five. We note that the number counts in general fields used for the comparison in \citet{Umehata2018} are derived from the best-fitting of combined data from various surveys \citep{Karim2013, Simpson2015, Aravena2016, Hatsukade2016, Oteo2016}, which is also shown in the right panel of Fig. \ref{fig: nc}. We notice that the best-fitting curve used in \citet{Umehata2018} is lower than the cumulative number counts in \citet{Gomez2022} and \citet{Chen2023a} which were used for our comparison. We compare our number counts with the fitting curve of general fields given in \citet{Umehata2018} and obtain an overdensity of 2.08 in the entire Spiderweb field, which is slightly higher to the result of $\sim$1.8 obtained from the comparison with \citet{Chen2023a} and \citet{Gomez2022}. We also list these results for each field compared to the field presented in \citet{Umehata2018}, in the last row of ALMA overdensity in Table \ref{tab: nc}.  
As previously mentioned, we also restrict our analysis to sources (\textgreater5$\sigma$) with purity close to one to calculate their cumulative number counts. This results in an overdensity of 1.7 at the lowest flux bin compared to field, consistent with those from the entire main catalog within 10\% uncertainties.

We note that the Spiderweb field was also covered by previous surveys observed with single-dish telescopes. The analysis of {\tt Herschel} 500\,$\mu$m sources confirmed an overdensity by a factor of two within 6\,cMpc around the central radio galaxy \citep{Rigby2014}, which is exactly the same as our results. \citet{Dannerbauer2014} reported 16 SMGs detected by APEX LABOCA over an area of $\sim$140\,arcmin$^2$ around the Spiderweb galaxy. The number density of LABOCA SMGs is up to four times that of general fields at 870\,$\mu$m down to  $S_{870\,\mu \rm m}=7$\,mJy. Similarly, 47 sources are detected by ASTE/AzTEC at 1.1\,mm in the Spiderweb field, over an area of  $\sim$210\,arcmin$^2$ \citep{Zeballos2018}. The 1.1\,mm AzTEC sources also show an overdensity by a factor of $\sim2$ compared to general fields. These results show a good agreement with our calculations, further confirming the overdensity nature of dust-obscured galaxies in the Spiderweb protocluster.

\subsection{Spatial correlation between different populations}

The spatial distributions of different member galaxies such as HAEs and DSFGs, cannot only reveal the local overdensities in protoclusters, but also help to depict the large scale structure over up to hundreds of comoving Mpc. Distinct sub-components (i.e., density peaks) and filamentary structures associated with protoclusters have been reported in several studies through different populations \citep{Matsuda2005, Casey2015, Cucciati2018, Shimakawa2018a, Zheng2021, Shi2021, Sun2024}.
Based on narrow-band imaging, the filamentary structures were identified over $\sim$10\,cMpc scale traced by HAEs, running across from the east to the south-west of the Spiderweb protocluster \citep{Koyama2013, Shimakawa2018b}. 
\citet{Jin2021} reported that CO emitters seems to trace the large scale structure extending to more than one hundred cMpc, suggestive of a super-protocluster or a filamentary structure linking to the Spiderweb field.

The surveys of the HAEs and CO emitters cover overall the footprint of our ALMA observation. In Figure \ref{fig: density map}, we show the positions of HAEs, CO emitters, and ALMA-detected DSFGs. We build the density maps of HAEs (in blue), DSFGs (in orange), and CO emitters (in green) based on the 5th neighbor analysis. The adopted method and the constructed HAE density map are identical to those of \citet{Shimakawa2018b}. It is clearly seen that the HAEs are highly concentrated around the Spiderweb galaxy. In contrast, the concentration of ALMA sources does not overlap with the HAE density peak centered on the radio galaxy, as reported by \citet{Dannerbauer2014}. The offset between these two populations was also found in two massive protoclusters at similar redshift \citep{Zhang2022} with a single-dish instrument. We can see that the CO emitters are roughly associated with the density peaks of the ALMA sources and are located near the filaments identified by the HAEs.

\begin{figure}
\includegraphics[width=\columnwidth]{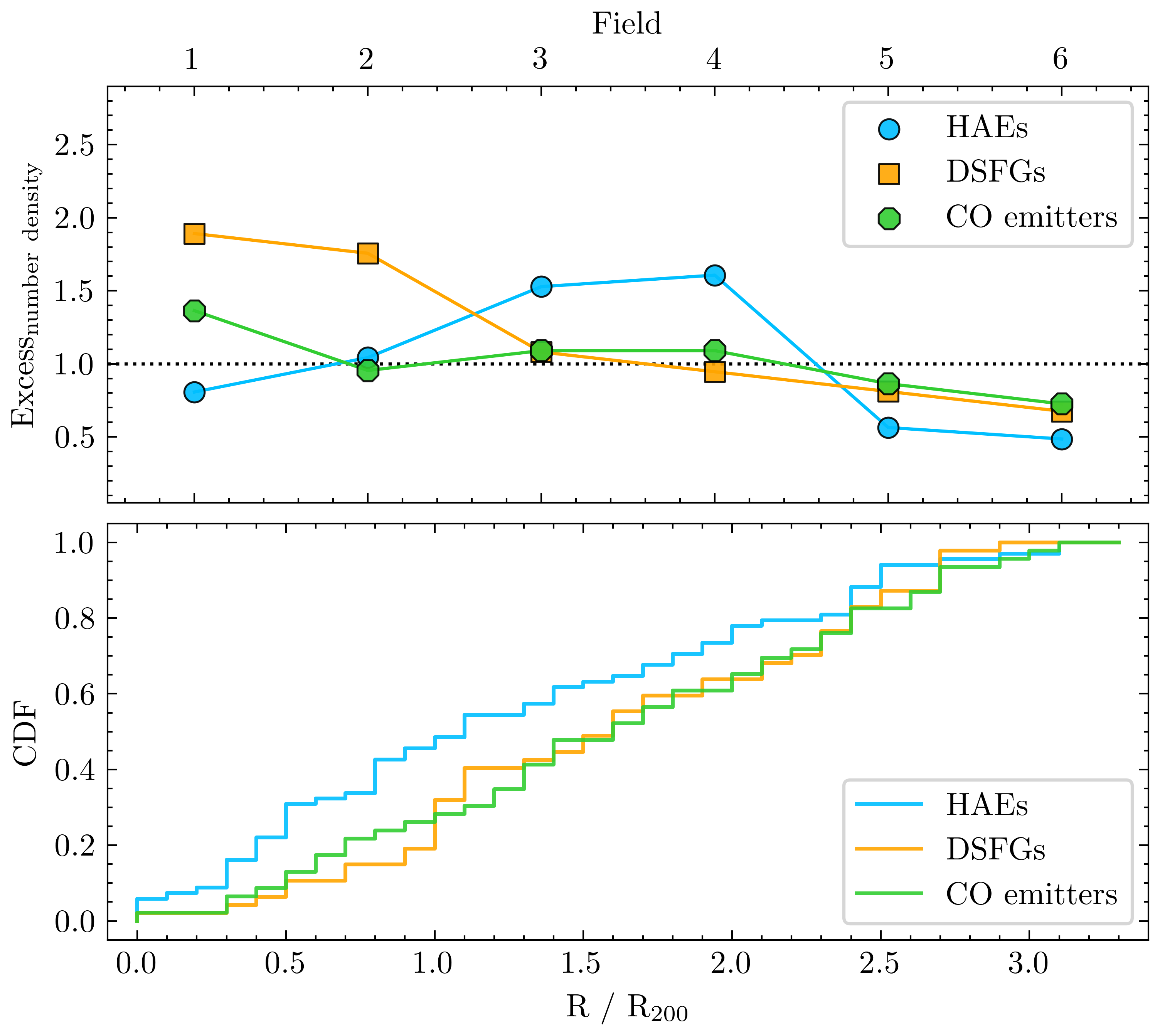}
  \caption{{\bf Upper}: Normalized excess of the number density of different source populations in each of the six fields. {\bf Lower}: The cumulative distribution function of the three populations as a function of radial distance towards the central Spiderweb galaxy. The HAEs, DSFGs and CO emitters are colored by blue, orange and Green in both panels. }
     \label{fig: pop_comparison}
\end{figure}

We quantitatively evaluated the overdensity factors of HAEs and CO emitters in each field and compared them with that of DSFGs, as the ALMA observations were equally divided into six fields from east to west. 
We note that the overall overdensity factors of DSFGs, HAEs and CO emitters vary a lot, and range from a factor of two to more than one order of magnitude.
We count the number of each population covered by our ALMA observations, and divide the number by the ALMA survey area of $\sim20\,{\rm arcmin}^2$ to obtain the average number density of each population over the entire ALMA footprint. We then calculate the surface density in each field with the same method and derive the excess of number density by normalizing them with the overall average number density. We show the results in the upper panel of Figure \ref{fig: pop_comparison}. It is clearly seen that the HAEs are highly concentrated in Fields 3 and 4, corresponding to the region surrounding the central Spiderweb galaxy. 
Both HAEs and DSFGs show significant excess in Field 1, exceeding their respective overall overdensities within the ALMA footprint. All three populations suggest lower overdensities in the western part, i.e., Fields 5 and 6, compared to the entire region of the protocluster.

In the lower panel of Fig. \ref{fig: pop_comparison}, we show the cumulative distribution function (CDF) of HAE, DSFGs and CO emitters as a function of distance to the central Spiderweb galaxy. Again, the HAEs are more concentrated towards the protocluster center than the other two populations. Approximately half of the HAEs are detected within the R$_{200}$, which is defined by \citet{Shimakawa2014} using both spectroscopically confirmed LAEs and HAEs. Only $\sim30$\,percent of DSFGs and CO emitters are found within the R$_{200}$, and these two populations show very similar spatial distributions. This may indicate that cold gas has accumulated along filaments on the outskirts of the protocluster, where galaxies were fed abundant gas and triggered with intense star formation before falling into the cluster center. 

\begin{table}[]
    \centering
    \caption{Results of 2D K-S test on the different source populations in the Spiderweb protocluster field.}
    \begin{tabular}{cccc}
    \toprule
    Populations & K-S statistic &  Probability & Significance \\
    \midrule
    HAE \& DSFG & 0.264 & 0.084 & 1.8$\sigma$ \\
    HAE \& CO  & 0.261 & 0.094 & 1.7$\sigma$ \\
    DSFG \& CO & 0.170 & 0.623 & 0.5$\sigma$ \\
    \bottomrule
    \end{tabular}
    \label{tab: ks_test}
\end{table}

Furthermore, we adopted a 2D Kolmogorov-Smirnov test \citep[K-S test;][]{Peacock1983, Fasano1987} to assess the correlation between HAEs, DSFGs, and CO emitters. In Table \ref{tab: ks_test}, we show the results of the K-S test on these three populations. By comparing HAEs with DSFGs, the probability of them being drawn from the same distribution is 0.084, similar to that between HAEs and CO emitters. However, given the low significance (\textless2$\sigma$) of K-S test between HAEs and the other two populations, we cannot determine that they are totally distinct populations. This is probably due to the fact that some HAEs are extremely massive star-forming galaxies, they are easily detected in dust and CO emissions and therefore overlap with the populations of DSFGs and CO emitters \citep{Gullberg2016, Emonts2016, Emonts2018, Dannerbauer2017}.

\citet{Zhang2022} reported that SMGs and HAEs are drawn from different spatial distributions in two protoclusters with higher significance from the K-S test. This may indicate that DSFGs and HAEs have stronger anticorrelation over a larger scale of $\sim$30\,cMpc and that their density peaks are representative of different local environments within protoclusters.  This trend is also suggested by the fraction of DSFGs/SMGs containing HAE counterparts. There are ten SMGs found to host HAE counterparts among a sample of 97 SMGs in the two MAMMOTH protoclusters \citep{Zhang2022}, resulting in a detection rate of around 10\,percent. When focusing on the spiderweb protocluster on a smaller scale, 10 of 47 DSFGs in our main catalog are HAEs within a matching radius of 1\,arcsec, making the detection rate of HAE counterparts to be 20\,percent (see Table \ref{table: main catalog}). We note that there is a significant difference between the spatial resolutions of the single-dish telescope used in \citet{Zhang2022} and ALMA in our work, while the consistency between our ALMA results and previous LABOCA results \citep{Dannerbauer2014} has already shown the authenticity of single-dish observations. More ALMA observations towards the Spiderweb field covering larger scale are required to confirm the relations between HAE and DSFG populations. 

\begin{figure}
\includegraphics[width=\columnwidth]{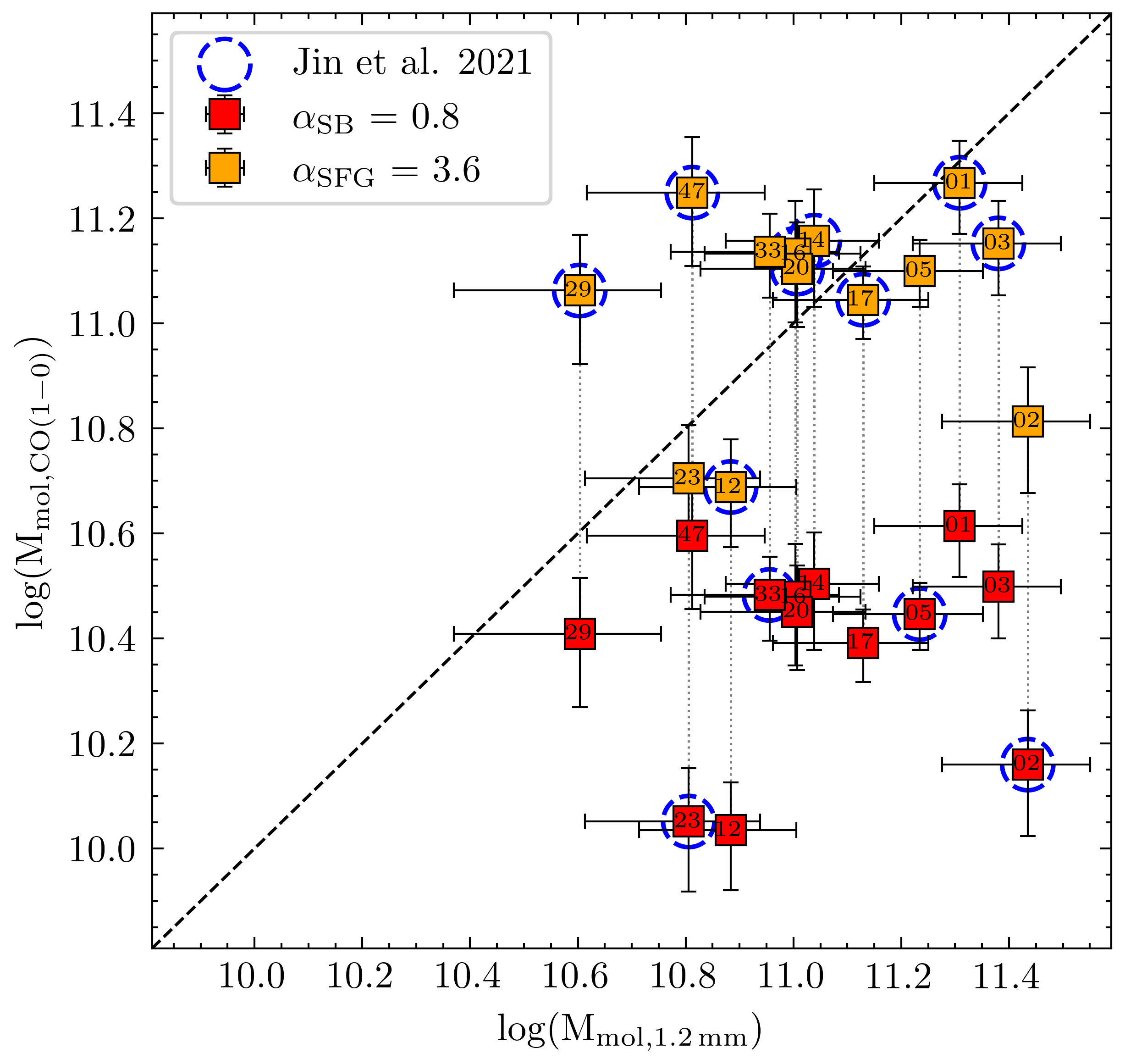}
\caption{Molecular gas content of the 13 ALMA sources with CO emitter counterparts, obtained from 1.2\,mm fluxes and CO(1-0) luminosities respectively. Different $\alpha_{\rm CO}$ values of 0.8 and $3.6\,{\rm M}_{\odot}\,({\rm K\,km\,s}^{-1}\,{\rm pc^2})^{-1}$ are adopted to convert the CO luminosities to molecular gas masses and marked as red and orange, respectively. The gas measurements by using the $\alpha_{\rm CO}$ assigned for each source following \citet{Jin2021} are shown in blue dashed lines.}
\label{fig: gas_mass}
\end{figure}

Interestingly, we find that ALMA-detected DSFGs and CO emitters have a rather high probability (62.3\%, see Table \ref{tab: ks_test}) coming from the same spatial distribution, consistent with the results of the CDF analysis shown in Fig. \ref{fig: pop_comparison}. This is reasonable as the molecular gas reservoirs are an indispensable fuel for star formation and DSFGs are dust obscured galaxies with intense star formation. The COALAS survey covers an area similar to that of our observations, and there are 43 ALMA sources in the main catalog located in the COALAS footprint. We note that the depth of the COALAS survey is not completely homogeneous over the whole map, see for details \citet{Jin2021} and \citet{Chen2024}.  We find that 13 ALMA sources are confirmed as CO emitters by cross-matching these two populations with a varied radius of $2-6$\,arcsec scaled by the ATCA synthesized beams \citep{Jin2021}. The matching results are the same as those using a fixed radius of 3\,arcsec adopted in \citet{Jin2021}. Interestingly, seven (6 robust, 1 tentative) out of the 13 CO emitters with ALMA counterparts are extended molecular gas reservoirs, indicating a fraction of extended gas reservoirs around 50\%, higher than that reported in \citet{Chen2024}. This further suggests that cluster galaxies with extended gas reservoirs are the results of efficient accretion of cold gas inflows along filaments, and starbursts fueled by sufficient gas are responsible for the growth of the stellar population in these galaxies \citep{Chen2024}.

This may raise these questions: If the ALMA sources and CO emitters are indeed correlated and indicative of intense star formation activities associated with significant gas inflows, why have we only detected 30\,percent (13/43) of ALMA sources with CO emissions?  We compare the depths of the ALMA and ATCA observations and estimate how sensitive they are in terms of molecular gas mass. \citet{Scoville2023} proposed \citep[please see also][]{Scoville2014,Scoville2016} that the gas mass can be determined from the dust luminosity at the Rayleigh–Jeans (RJ) tail due to its optically thin nature. By assuming a dust temperature of 25\,K, spectral index $\beta=2$, and a constant gas-to-dust-mass ratio of 105 \citep{Draine2011}, we derive the gas masses of our ALMA sources to be $2.0-27.3\times10^{10}\,{\rm M}_{\odot}$ based on the Gaussian fitting flux measurements. 
We note that the gas masses obtained above from dust luminosity are based on the assumption of normal star-forming galaxies \citep{Scoville2023}. \citet{Jin2021} divided their CO emitters into two regimes, and used a $\alpha_{\rm CO, SB}=0.8$ and $\alpha_{\rm CO,SFG}= 3.6$ (${\rm M}_{\odot}\,[{\rm K\,km\,s}^{-1}\,{\rm pc^2}]^{-1}$) to obtain the gas masses of starbursts and star-forming galaxies, respectively.

In Figure \ref{fig: gas_mass}, we compare the molecular gas masses obtained from 1.2\,mm fluxes and CO luminosities of the 13 ALMA sources with CO emitter counterparts. Following \citet{Jin2021}, we assign different $\alpha_{\rm CO}$ values for each source individually, and mark them with blue dashed circles. It is clearly seen that the gas masses obtained from dust and CO emissions do not agree with each other. Moreover, we use a single $\alpha_{\rm CO}$ value of $0.8\,{\rm M}_{\odot}\,({\rm K\,km\,s}^{-1}\,{\rm pc^2})^{-1}$ for all the 13 ALMA sources and repeat this process with a higher $\alpha_{\rm CO}$ value of $3.6\,{\rm M}_{\odot}\,({\rm K\,km\,s}^{-1}\,{\rm pc^2})^{-1}$. We can see from Figure \ref{fig: gas_mass} that the gas measurements from dust emission are significantly different with the lower $\alpha_{\rm CO}$ conversions, while comparable with the gas masses derived from higher $\alpha_{\rm CO}$ with large scatters. Thus, we conclude that the reliability of molecuar gas masses derived through dust measurements have to be seen with caution. 
By adopting an $\alpha_{\rm CO}$ of $3.6\,{\rm M}_{\odot}\,({\rm K\,km\,s}^{-1}\,{\rm pc^2})^{-1}$, we estimate the CO emitters with a lower limit of gas mass to be $5.0\times10^{10}\,{\rm M}_{\odot}$ in the COALAS survey, corresponding to an observing flux of 0.4\,mJy at 1.2\,mm.  This means that about half of ALMA sources have gas reservoirs below the ATCA detection limit.

\begin{figure}
\includegraphics[width=\columnwidth]{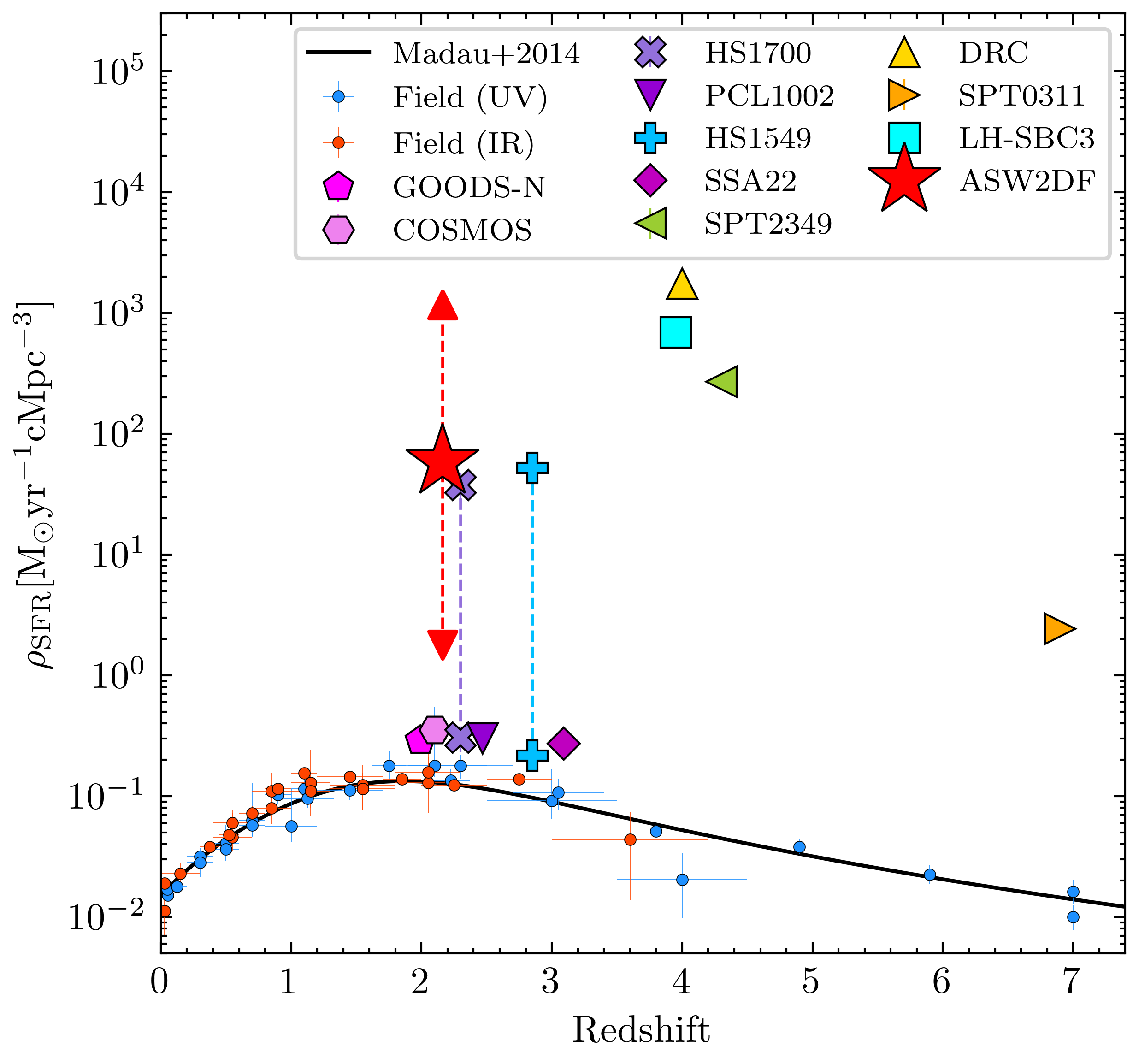}
\caption{Cosmic star formation rate density. The  measurements for field galaxies and the best-fitting function are shown in dots (blue from UV and red for IR) and black curve \citep{Madau2014}. We also show the results of different protoclusters including GOODS-N protocluster, COSMOS protocluster and SSA22 \citep{Casey2016}, PCL1002 \citep{Champagne2021}, HS1700+64 and HS1549+19 \citep{Lacaille2019} with both overall and core estimations (linked by the dashed vertical lines with the same color), SPT2349$-$56 \citep{Hill2020}, Distant Red Core \citep[DRC,][]{Oteo2018}, SPT0311$-$58 \citep{Wang2021}, and LH-SBC3 \citep{Zhou2024}. The SFR density calculated in this work is shown as an red star, based on the volume of a 2\,Mpc sphere (core-like) adopted in \citet{Dannerbauer2014}. The volumes are also estimated based on group size ($\sim$400\,kpc) and the whole ALMA field of view ($48\times120\,{\rm Mpc}^3$), as shown in upward and downward arrows, respectively.}
\label{fig: sfrd}
\end{figure}

In addition, we note that the CO emitters are detected within a quite large velocity range, corresponding to $\sim$120\,cMpc along the line of sight \citep{Jin2021}, much larger than the spatial coverage of both the ALMA and ATCA observations which have a scale of $\sim6\times9$\,cMpc$^2$. The 2D K-S test are solely based on the sky positions and thus cannot determine the correlation between DSFGs and CO emitters in redshift space.  The upcoming analysis on the JWST data \citep[GO 1572,][]{Dannerbauer2021} and more spectroscopic observations will help to confirm the membership of these ALMA sources, and unveil the nature of these dusty sources in relation to the CO emitters and surrounding large scale structures.

In Fig. \ref{fig: density map}, we also show the seven LABOCA sources \citep[DKB01, DKB02, DKB03, DKB07, DKB12, DKB13 and DKB15; see][]{Dannerbauer2014} and five AzTEC sources \citep{Zeballos2018} within the coverage of our ALMA observations. All of them are revolved by our ALMA observations and located at the eastern part of the Spiderweb protocluster. In particular, there are three LABOCA sources and two AzTEC sources in Field 1, making this region an extreme concentration of DSFGs over the entire Spiderweb field. This also matches well with previous results based on the field-to-field number counts and the CDF analysis. The similar distributions of LABOCA/AzTEC sources and ALMA sources verify that DSFGs revealed by ALMA are indeed concentrated in the eastern part of the Spiderweb field, without affected by the observing differences between six fields (see Table \ref{tab: obs}).

\citet{Dannerbauer2014} reported, based on the coarse spatial resolution of LABOCA, that the spatial distribution of SMG overdensity seems to be similar to the eastern filaments. Overall, we can see from Fig. \ref{fig: density map}, that both the DSFGs and CO emitters are concentrated close to the filaments traced by HAEs, even though they are likely drawn from different spatial distributions compared to that of HAEs. This is consistent with the basic concept that galaxy clusters assemble their mass by accreting gas and galaxies from the cosmic web through filaments as highways \citep{Zel'dovich1970, Rost2021}. The abundant gas supplies were cooling down and sustained around the filaments by accretion shocks \citep{Rost2021, Rost2024}, and galaxies close to the filaments on the outskirts of the protocluster were fueled by cold gas accretion \citep{Dekel2009, Daddi2021}. On the other hand, the accumulation of gas materials and shocks could trigger starbursts in these DSFGs through more frequent galaxy mergers and interactions, which are closely related to the local environments \citep{Liu2023, Naufal2023}.  To this end, a panoramic mapping of three-dimensional Ly$\alpha$ emission is crucial to reveal the underlying gas reservoirs and cold streams not only in the center of the protocluster, but also on the outskirts extending to several Mpc \citep{Umehata2019, Martin2023}. 

We calculate the cosmic SFR density of the ALMA sources over a survey area of $\sim$20\,arcmin$^2$, corresponding to 48\,cMpc$^2$. We firstly convert the fluxes of these ALMA sources into fluxes at 850\,$\mu$m according to the ALESS SED \citep{da_Cunha2015}, then derive the SFR for each source based on the linear relation from \citet{Cowie2017}. The derived SFR density depends significantly on the assumed volume. If conservatively, we adopt a ligh-of-sight distance of 120\,cMpc from \citet{Jin2021}, in which the CO emitters are assumed to trace a super structure, we obtain an SFR density to be $1.78\pm0.35\,{\rm M}_\odot\,\rm{yr}^{-1}\rm{cMpc}^{-3}$, being around 14 times that of the average cosmic SFR density \citep{Madau2014} and well above the predictions of simulations \citep{Bassini2020}. However, if considering a 2\,Mpc sphere as adopted in \citet{Dannerbauer2014}, which is representative of a core-like volume, we obtain a SFR density to be $58.21\pm11.41\,{\rm M}_\odot\,\rm{yr}^{-1}\rm{cMpc}^{-3}$. Moreover, within a smaller group size of $\sim$400\,kpc sphere (see Sec. \ref{sec: discuss3}), we would obtain a much higher SFR density of $1171.25\pm233.20\,{\rm M}_\odot\,\rm{yr}^{-1}\rm{cMpc}^{-3}$. We show our results in Figure \ref{fig: sfrd}, together with those from protoclusters at different scales and fields from the literature. The SFR density of our ALMA sources in a core-like 2\,Mpc sphere agrees very well with that of \citet{Dannerbauer2014}, and are comparable to the cores of other protoclusters at similar redshift.  When focusing on the group size, the estimated SFR density will be even much higher than coeval protoclusters, while comparable to the starburst group LH-SBC3 at $z=3.95$ \citep{Zhou2024}, DRC \citep{Oteo2018} and STP2349$-$56 \citep{Hill2020}. We emphasize that due to the uncertainty of the calculations of the cosmic volume, one should take the cosmic SFR density with caution. 

\subsection{DSFGs related to the evolution of the Spiderweb protocluster}
\label{sec: discuss3}

The high concentration of DSFGs in a small region might indicate the existence of a protocluster core or galaxy groups with intense starburst activities. There are 14 (23) sources over a physical scale of 130 (300)\,kpc confirmed by ALMA in the core of the protocluster SPT2349$-$56 at $z=4.3$ \citep{Miller2018, Hill2020}. An extreme protocluster core, DRC, was also discovered with at least 10 ALMA sources over $\sim$300\,kpc at $z=4.0$ \citep{Oteo2018, Long2020}. \citet{gomez2019} reported two {\tt Herschel} protoclusters, HELAISS02 at $z=2.17$ and HXMM20 at $z=2.6$, confirmed by ALMA with four and five detections over a similar scale. Most recently, the first results of the NICE large program showed that a starburst galaxy group at $z=3.95$ with four members was discovered within a diameter of 180\,kpc \citep{Zhou2024}. 

These results motivate us to estimate the excess of the DSFGs over a smaller scale and examine their local environment. In Fig. \ref{fig: density map}, we can see that there are two distinct concentrations of ALMA-detected DSFGs around DKB01/DKB03 and DKB02 respectively. There are eight (six) DSFGs around DKB01/DKB03 (DKB02) over a scale of 400\,kpc, suggesting an excess by a factor of \textgreater12 compared to that in general fields. We tentatively regard these concentrations as two galaxy groups, SW-GE1 and SW-GE2, with intense star formation activities. We calculate the total SFR of these two groups to be $2688\pm385$ and $806\pm121\,{\rm M}_\odot\,\rm{yr}^{-1}$.  The estimated SFRs of these groups are comparable with that of the central Spiderweb complex \citep{Shimakawa2018b}. 
However, when focusing on the traditional HAE overdensity peaks centered on the Spiderweb galaxy (ASW2DF.17), we find only two ALMA sources within a comparable physical scale. Similarly, the SMG DKB12 has been confirmed as a galaxy group containing at least four HAE members within 60\,kpc \citep{Dannerbauer2014, Shimakawa2018b}, which is even smaller than the scale of 400\,kpc adopted here. Only one ALMA source (ASW2DF.01) is related to these HAEs. This is consistent with previous findings, which show that the clustering of ALMA-detected DSFGs is clearly deviated from HAE density peaks.

We note that three out of eight ALMA-detected DSFGs (ID: 02, 03, 33) in group SW-GE1 have been spectroscopically confirmed by either CO(1-0) and/or H$\alpha$ emission \citep{Jin2021, Perez2023}. Moreover, two previously reported HAEs but undetected by ALMA are located in this region \citep{Perez2023}, and one of them is also a CO emitter \citep{Jin2021}. 
These five confirmed members have already proven that SW-GE1 contains an extreme overdensity of DSFGs of at least 10, and even more when considering their volume density. 
Three additional ALMA sources (ID: 08, 09, 13) at \textgreater5$\sigma$ are identified in this group, with two exhibiting clear $K_{\rm s}$ counterparts \citep{Shimakawa2018b}, suggesting they are likely group members.
Interestingly, the sources ASW2DF.03 and ASW2DF.33, which have been previously confirmed by \citet{Dannerbauer2014} and dubbed as DKB01b and DKB03, are recently reported to host extended gas reservoirs up to $\sim70$\,kpc \citep{Chen2024}. 
It is clearly seen that the robust candidates for extended gas reservoirs are mostly located in CO density peaks along the filaments (see Fig. \ref{fig: density map}).
These results support the idea that the gas fuels are preferentially located in denser local environments and suspected to be the results of effective accretion of cold streams along filaments, then various mechanisms like gas truncation, galaxy mergers/interactions are attributed to the causes of the extended gas reservoirs and the growth of proto-ICM \citep{Chen2024}. 

As for the other group SW-GE2 with six ALMA sources (ID: 04, 19, 24, 37, 45, 47), none of them are identified as HAEs nor CO emitters. Only ASW2DF.04, the brightest one with a secure counterpart detected with the Karl G. Jansky Very Large Array (VLA) \citep{Dannerbauer2014}, has recently been confirmed as a member galaxy at $z=2.15$ (Naufal et al., in prep.). We exclude the source ASW2DF.37 which overlaps with an AGN at $z=1.512$ \citep[X9, see][]{Pentericci2002, Croft2005}. 
Furthermore, only half of the six ALMA sources have a S/N higher than five, making the authenticity of this galaxy group less certain.
This situation is similar to the one that has been reported in \citet{Chen2023b}, in which the authors found that an assumed overdensity of dusty galaxies is actually composed by seven galaxies distributed at a wide redshift range rather than being a real association. 
Nevertheless, the membership of the other four ALMA sources and their potential physical relation to the large scale structure should be investigated with further spectroscopic observations.


It is evident that DSFGs are grouped and host intense star formation activities on the outskirts of the Spiderweb protocluster up to $\sim3\times{\rm R}_{200}$. Contrarily, the protocluster center shows a relative lack of DSFGs and is believed to be nearly virialized \citep{Dannerbauer2014, Shimakawa2014, Jin2021}. The member HAEs located in the protocluster center are at the high-mass end, likely in a post-starburst phase and with higher AGN fraction \citep{Shimakawa2018b, Shimakawa2024, Perez2023}. This is consistent with the scenario proposed by \citet{Shimakawa2018b}, in which the Spiderweb protocluster is experiencing a ``maturing phase''. In this scenario, the red, passive galaxies have emerged in the protocluster center \citep{Kurk2004b, Kodama2007, Zirm2008, Doherty2010, Tanaka2010, Tanaka2013}, while massive HAEs are maturing galaxies undergoing a rapid transition to quenched populations, with active AGN feedback responsible for suppressing the star formation activities in these HAEs \citep{Saro2009, Shimakawa2024, Perez2024}. Some of them host post-starburst features and thus are suspected to be starbursts previously in their evolution \citep{Shimakawa2018b}. For the ALMA-detected DSFGs located on the outskirts, they are still forming stars intensely because of efficient gas accretion and frequent interaction/merger events. Furthermore, these DSFGs may have assembled most of their masses before accreting into the potential well in the protocluster center \citep{Smail2014, Dannerbauer2014}. A similar distribution has been reported by \citet{Shi2024}, who found a strong anti-correlation between star formation activities and the density profile in a massive protocluster at the same epoch. This is consistent with the inside-out growth scenario proposed in simulations, that protoclusters at $z\sim1.5-5$ were experiencing rapid star formation over an extended scale of $10-20$\,Mpc \citep{Chiang2017}, and most of the galaxies are rapidly evolving inside the collapsing ﬁlaments instead of the protocluster core \citep{Remus2023}. 

\begin{figure}[h!]
\includegraphics[width=\columnwidth]{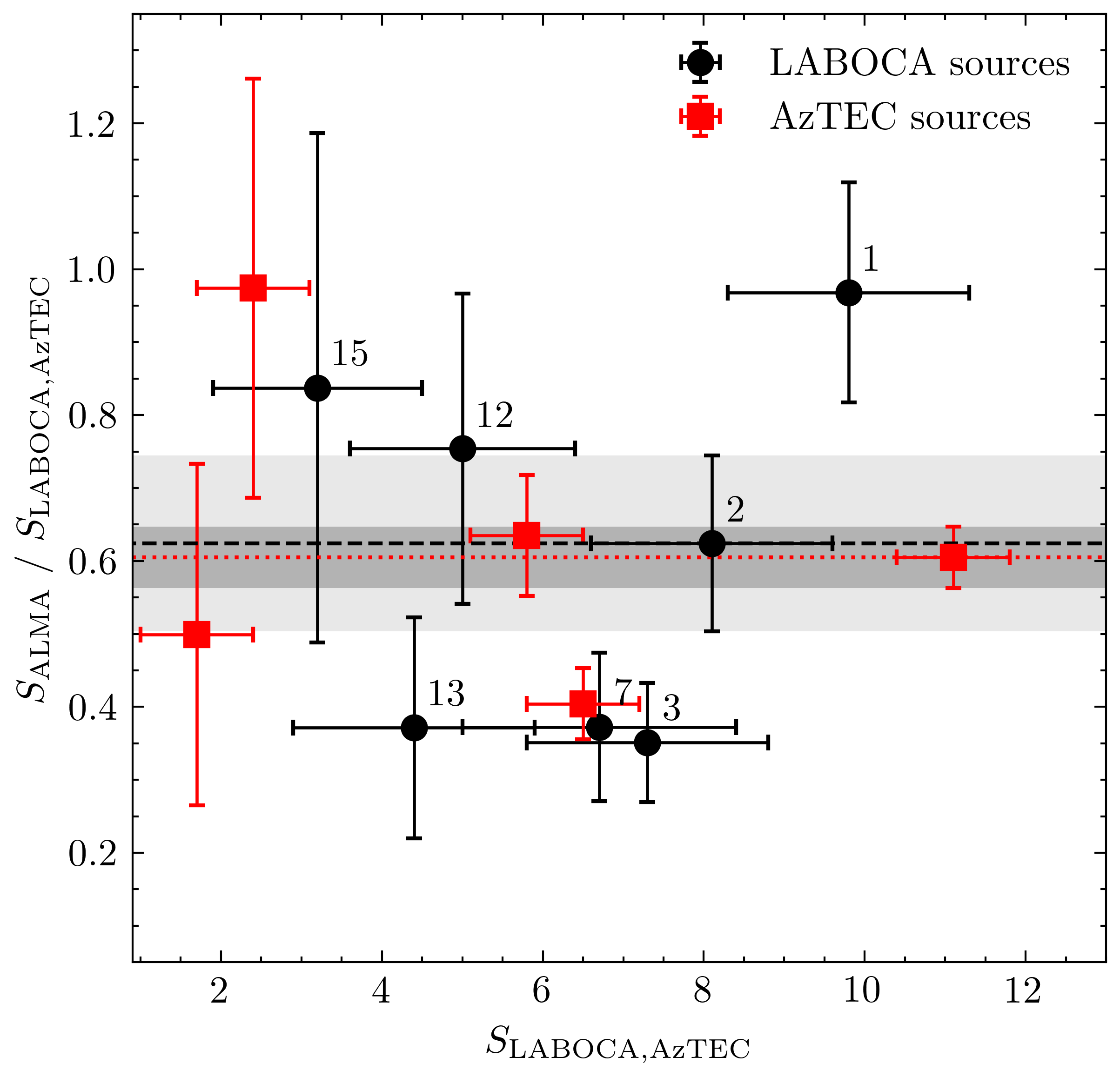}
  \caption{The flux ratio between the ALMA and single-dish sources as a function of the original single-dish flux. The ALMA flux are scaled to 870\,$\mu$m and 1.1\,mm to match the LABOCA (black) and AzTEC (red) sources. The numbers adjacent to the black dots correspond to the IDs of DKB sources as identified in \citet{Dannerbauer2014}. The dashed and dotted lines show the median of the flux ratios, together with the scatters marked by the shaded regions. }
     \label{fig: flux_ratio}
\end{figure}

On the other hand, the Spiderweb protocluster is considered to be the result of the merger of two massive halos based on the double-peaked velocity distribution of the spectroscopically confirmed members \citep{Pentericci2000, Kuiper2011}. Simulations have indicated that a significant fraction of the cluster mass was assembled through the accretion of galaxy groups \citep{McGee2009} and several merger events could occur in a protocluster during its virialization \citep{Kimmig2023}. We might be witnessing a late phase of a major merger between two galaxy groups within the protocluster center, where the HAEs show an enhanced AGN fraction, and some of them are hosting post-starburst features \citep{Shimakawa2018b, Tozzi2022a}. The anisotropic nature of the merged subhalos could be responsible for the asymmetric proto-ICM discovered recently \citep{Mascolo2023, Lepore2024}. We thus speculate that the galaxy groups SW-GE1 and SW-GE2 will eventually fall into the potential well of the protocluster center and another merger will be triggered. Using the previously estimated SFRs and gas masses, we could have a rough idea of the depletion time of these ALMA sources, despite the inherent degeneracy between the two quantities based only on dust measurements. Without newly replenished gas inflows, the depletion time is estimated to be $340\pm48$ Myr, i.e., these ALMA-detected DSFGs will consume up their gas reservoirs and become passive at $z\sim1.9$. 

\subsection{Comparison between ALMA and single-dish observations }
\label{sec: diss_single_dish}
The preceding single-dish observations conducted on the Spiderweb protocluster provide an opportunity to match and compare the ALMA and LABOCA/AzTEC detections. All the seven LABOCA sources are detected by ALMA with at least one ALMA counterpart. Four of them are recovered with multiple (two or more) ALMA detections, indicating a multiplicity fraction of $\sim$57\%. The multiplicity fraction is higher than that in general fields \citep[\textless50\%, see][]{Stach2018, Simpson2020}. All the brightest LABOCA sources DKB01, DKB02 and DKB03 show multiplicity, consistent with the trend that the bright single-dish sources are more likely to consist of multiple ALMA sources \citep{Hodge2013, Stach2018, Simpson2020}. Furthermore, all five AzTEC sources within the ASW$^2$DF survey coverage are detected by ALMA, and four of them are recovered with more than one ALMA source. 
The 80\% multiplicity of AzTEC sources is even higher than that in the protocluster SSA22 \citep[$\sim$63\%, see][]{Umehata2018}, under the same instrument beam size of FWHM=30\,arcsec. The higher multiplicity of single-dish detections revealed by ALMA supports the overdensity nature of DSFG population in the Spiderweb protocluster.

We compare the flux measurements between ALMA and LABOCA/AzTEC sources to check how much of the single-dish flux are recovered by the ALMA observations. We sum up the fluxes of ALMA sources located within each LABOCA/AzTEC beam. 
The fluxes of the single-dish sources are scaled to 1.2\,mm by using the same method described in Sec. \ref{sec: nc}, with $S_{\rm 1.2\,mm}/S_{\rm 870\,\mu m}$ of 0.40 and $S_{\rm 1.2\,mm}/S_{\rm 1.1\,mm}$ of 0.77. Fig. \ref{fig: flux_ratio} shows the ratio of the summed ALMA flux compared to the single-dish flux. In most cases, the single-dish flux is not fully reproduced by the ALMA observations. The median ratios for the LABOCA sources and AzTEC sources are $0.62\pm0.12$ and $0.60\pm0.04$, which agree well with the results of \citet{Umehata2018} within uncertainties. Similarly, \citet{Hill2020} found that the integrated flux of the ALMA sources at 850\,$\mu$m only account for $55\pm5$\,percent of the integrated LABOCA flux at 870\,$\mu$m in the core region of the protocluster SPT2349$-$56, meaning that there are still many galaxies under the detection limit due to the low SFR or spatial resolution.

There are several causes which could explain the missed flux by the ALMA sources. First of all, the flux of LABOCA sources adopted in \citet{Dannerbauer2014} is the observed flux, which could be boosted by the Eddington bias and confusion noise.  Secondly, the fluxes of ALMA sources are recovered better with increasing beam sizes, and more flux may be lost with Briggs weighting compared to natural weighting \citep{Oteo2017}.  Our observations could recover most of the source fluxes in principle, but may still miss some fluxes due to the weighing scheme and the high resolution. Especially considering that there are a number of DSFGs showing clear extended features (i.e., cloud-like), whose emission cannot be fully covered by the aperture we used for flux measurement. This is also indicated by the big differences (\textgreater50\%) between the flux measurements from aperture and Gaussian fitting, such as sources with ASW2DF IDs of 08, 33, 35, 41 and 44 (see Figure \ref{fig: gallery_main} and \ref{fig: gallery_supp}). Furthermore, there could be fainter dusty galaxies undetected by ALMA that contribute to the single-dish flux, as described in \citet{Umehata2018}. Finally, the observing frequency of the ALMA observations is not exactly the same as that of single-dish surveys \citep{Simpson2015, Simpson2020}, the flux conversion factor could overestimate the single-dish flux and then lower the flux ratios. For example, \citet{Gomez2022} adopted a flux conversion factor of 0.44 for $S_{\rm 1.1\,mm}/S_{\rm 870\,\mu m}$ and 1.36 for $S_{\rm 1.1\,mm}/S_{\rm 1.2\,mm}$, which will decrease the single-dish flux by up to 30\,percent.

\section{Summary}
\label{sec: summary}
We present the first results of the ASW$^2$DF survey targeting the prominent Spiderweb protocluster at $z=2.16$. The observations were carried out in ALMA band 6. The data were reduced and cleaned into contiguous mosaic maps in six fields, with a total survey area of around 20\,arcmin$^2$. We tested the ``individual'' and ``fixed'' methods on the rms noise estimation (see Sec. \ref{sec: noise}). The ``fixed'' method is more conservative and adopted to estimate the typical 1$\sigma$ noise in each field, being in a range of $40.3-57.1\,\mu$Jy. Below we summarize our main results in this paper:

\begin{enumerate}
    \item We used three publicly available codes (SEP, AEGEAN and {\sc SoFiA}) to extract sources in the ALMA maps at S/N\textgreater4. With the conservative ``fixed'' method, the three codes give consistent results and a sample of 47 detections was obtained as the main catalog, with a high purity of \textgreater90\,percent. Additionally, we performed a careful visual inspection on the mosaic maps to identify the ALMA sources independently. A supplementary catalog is provided to contain 21 sources identified by visual inspection or detected by any of the three codes with the ``individual'' method, which cannot be detected by the ``fixed'' method at \textgreater4$\sigma$. 

    \item Based on the main catalog, we constructed the differential and cumulative number counts down to 0.25\,mJy at 1.2\,mm. 
    By comparing our results with those of general fields and public protoclusters, we find that the number counts of our ALMA sources are around two times that in general fields, and consistent with previous results with single-dish instruments. In addition, the number counts in Fields 1 to 3 are higher than the overall estimation, while Fields 4 to 6 show slightly lower number counts.
     
    \item The comparison between the spatial distributions of our ALMA sources with HAEs and CO emitters in the same field, reveals that the DSFGs detected by ALMA are preferentially located at the outskirts of the Spiderweb protocluster, and likely drawn from the same distribution as that of CO emitters. This is consistent with simulations that the galaxies and gas were falling into the cluster through filaments, where the starbursts are triggered by abundant gas accretion and frequent interactions/mergers due to the accretion shocks \citep{Zel'dovich1970, Rost2021}. 
    We find that 30\,percent of ALMA sources located in the COALAS footprint are detected through CO(1-0) observations of our COALAS large program. The main reason for the non-detections could be the shallow (and not completely uniform) depth of the CO(1-0) observations. The SFR density of our ALMA sources is consistent with previous results \citep{Dannerbauer2014} after accounting the difference in volume. Using dust measurements in order to derive molecular gas masses should be considered with caution.
    
    \item Two extreme concentrations of ALMA sources around DKB01/DKB02 and DKB03 are located in the eastern part of the protocluster, with an excess of 12 times that of general fields, denoted as SW-GE1 and SW-GE2. This supports the scenario proposed by \citet{Shimakawa2018b} that the Spiderweb protocluster is in a maturing phase. Here, some massive HAEs, exhibiting post-starburst features, are located in the nearly virialized protocluster center. In contrast, the DSFGs on the outskirts are undergoing starbursts due to efficient gas accretion before their accretion into the protocluster center. These two newly identified groups are likely to eventually fall into the protocluster center and trigger another merger event.

    \item Comparing our ALMA maps with previous single-dish observations, we found that the multiplicity fraction is higher for the single-dish SMGs in the Spiderweb protocluster than that in general fields. This could also explain the overdense nature of DSFGs in the protocluster. The flux comparison indicates that the ALMA recovered around 60\,percent of the single-dish fluxes. The potential reasons for the flux loss could be: flux boosting in single-dish observations, missing flux issue due to high resolution of ALMA, extended and/or faint DSFGs which are larger than the apertures used for flux measurement, and uncertainty of flux conversion between observed wavelengths. 
\end{enumerate}

\begin{acknowledgements} 
This paper makes use of the following ALMA data: ADS/JAO.ALMA\#2021.1.00435.S ALMA is a partnership of ESO (representing its member states), NSF (USA) and NINS (Japan), together with NRC (Canada), MOST and ASIAA (Taiwan), and KASI (Republic of Korea), in cooperation with the Republic of Chile. The Joint ALMA Observatory is operated by ESO, AUI/NRAO and NAOJ. The National Radio Astronomy Observatory is a facility of the National Science Foundation operated under cooperative agreement by Associated Universities, Inc. The Australia Telescope Compact Array is part of the Australia Telescope National Facility (grid.421683.a) which is funded by the Australian Government for operation as a National Facility managed by CSIRO. We acknowledge the Gomeroi people as the traditional owners of the Observatory site. HD, YZ, JMPM and ZC acknowledges ﬁnancial supports from the Agencia Estatal de Investigación del Ministerio de Ciencia e Innovación (AEIMCINN) under grant (La evolución de los cúmulos de galaxias desde el amanecer hasta el mediodía cósmico) with reference (PID2019105776GB-I00/DOI:10.13039/501100011033). In addition, HD, YZ and JMPM acknowledge support from the Agencia Estatal de Investigación del Ministerio de Ciencia, Innovación y Universidades (MCIU/AEI) under grant (Construcción de cúmulos de galaxias en formación a través de la formación estelar oscurecida por el polvo) and the European Regional Development Fund (ERDF) with reference (PID2022-143243NB-I00/10.13039/501100011033). YZ acknowledges the support from the China Scholarship Council (202206340048), and the National Science Foundation of Jiangsu Province (BK20231106). JMPM acknowledges funding from the European Union’s Horizon research and innovation programme under the Marie Skłodowska-Curie grant agreement No 101106626. XZZ acknowledges the support from the National Science Foundation of China (12233005 and 12073078). CDE acknowledges funding from the MCIN/AEI (Spain) and the ``NextGenerationEU''/PRTR (European Union) through the Juan de la Cierva-Formación program (FJC2021-047307-I). YK, RS, and TK acknowledge support from JSPS KAKENHI Grant Number JP23H01219.
\end{acknowledgements}

\bibliographystyle{aa}
\bibliography{aa.bib}

\begin{appendix} 

\onecolumn
\section{Additional material}
\label{sec: appendix}

\begin{figure}[ht!]
\centering
\includegraphics[width=0.9\textwidth]{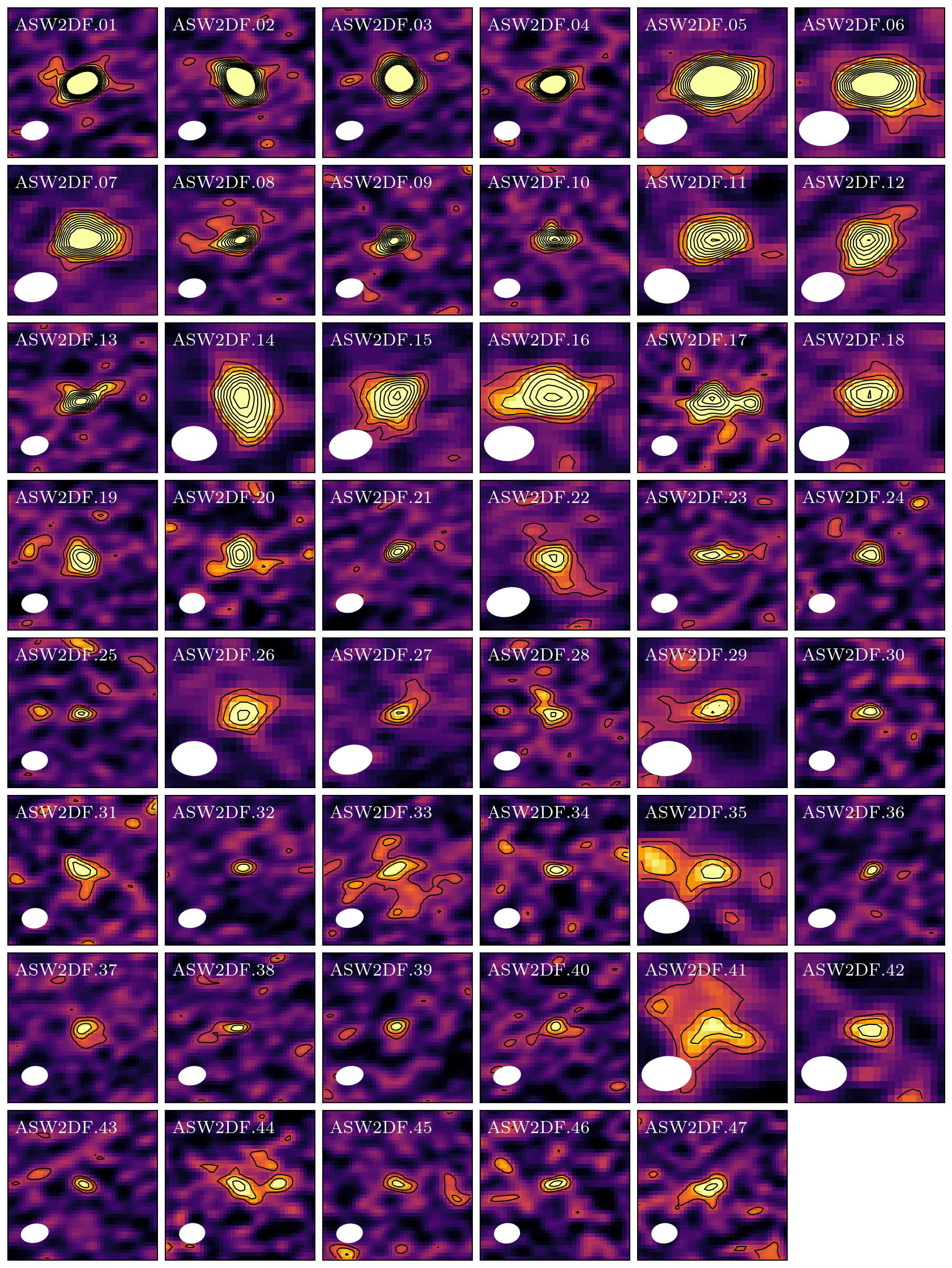}
  \caption{Gallery of the ALMA 1.2\,mm sources in Spiderweb field. The 47 sources are from the main catalog and labeled with the ID in the upper left corner of each cutout. The contours start with 2$\sigma$ in steps of 1$\sigma$ noise in each field, respectively. The synthesized beam for each source is shown in the white ellipse. The image size is $3^{\prime\prime}\times3^{\prime\prime}$ with the north on the up and the east on the left.}
     \label{fig: gallery_main}
\end{figure}

\begin{figure}
\centering
\includegraphics[width=0.9\textwidth]{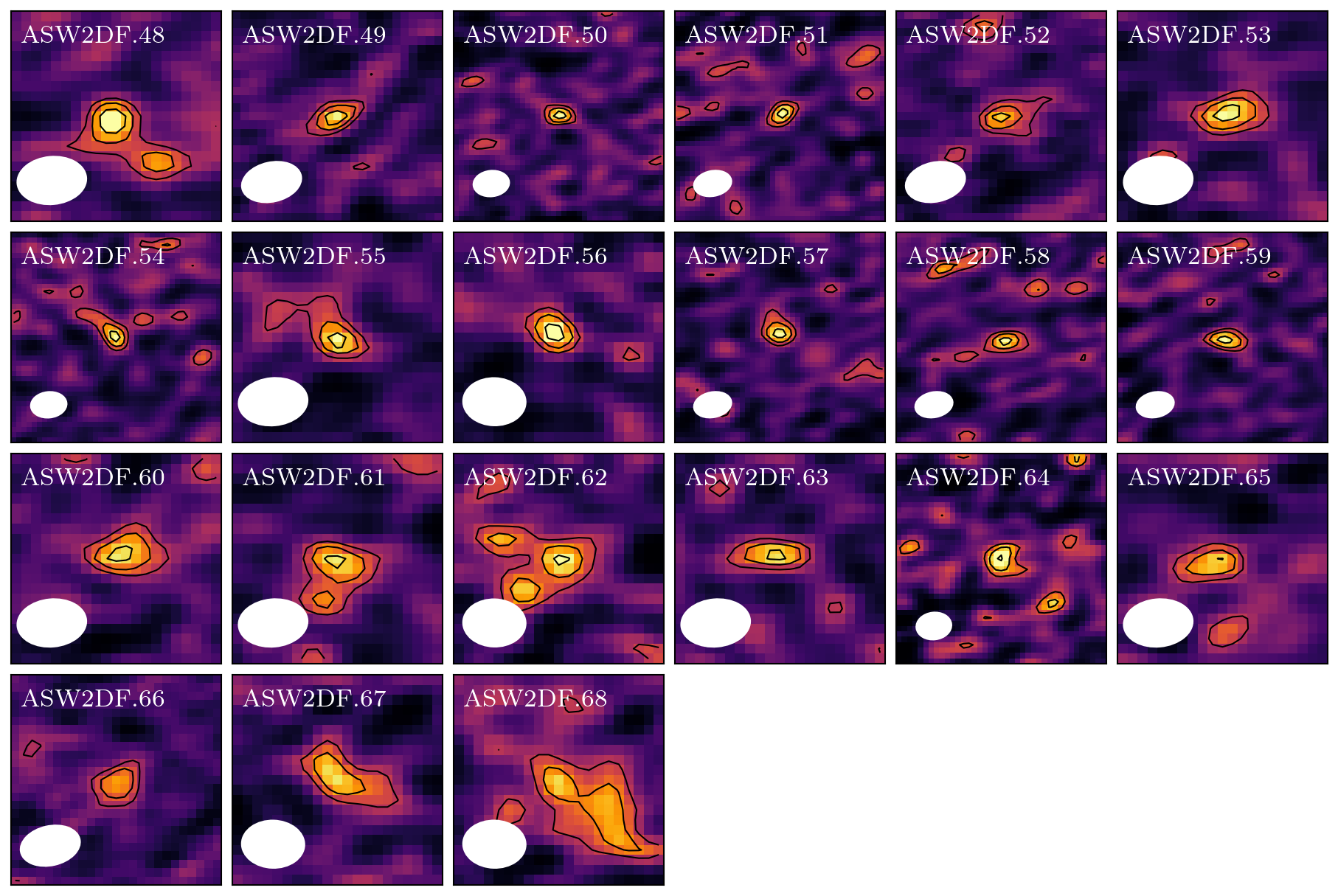}
  \caption{Gallery of the ALMA 1.2\,mm tentative sources in Spiderweb field. The 21 sources are from the supplementary catalog and plotted in the same manner as Fig. \ref{fig: gallery_main}.}
     \label{fig: gallery_supp}
\end{figure}

\begin{table*}
\centering
\caption{Summary of the observations of the ASW$^2$DF survey.} 
\label{tab: obs}
\begin{tabular}{cccccccccccc}
    \toprule
    Field   
    &  EB 
    & Date   
    & Ant. 
    & Baseline 
    & Resolution
    & $\tau_{\rm target}$ 
    &  PWV 
    &  EF 
    &  Synthesized beam 
    & PA
    & $\sigma$ \\
    &  &  &  & (m) & (arcsec) & (min) & (mm)  &   &  & (deg) &($\mu$Jy) \\
    \midrule
      & 1 & 31 Dec 2021 & 42 & 14.9$-$976.6 & 0.4$-$5.0 & 40.57  & 1.49  & 1.10 &  &  &  \\
     1 & 2 & 02 Jan 2022 & 43 & 14.9$-$976.6 & 0.4$-$5.2 & 40.55  & 0.96  & 1.50 & 0\farcs51\,$\times$\,0\farcs33 & $-$78.6 & 46.2 \\
      & 3 & 02 Jan 2022 & 44 & 14.9$-$976.6 & 0.4$-$5.1 & 40.57  & 0.79  & 1.50 &  &  &  \\

      & 1 & 03 Jan 2022 & 42 & 14.9$-$783.1 & 0.4$-$5.3 & 40.53  & 1.33  & 1.31 &  &  &  \\
     2 & 2 & 03 Jan 2022 & 44 & 14.9$-$976.6 & 0.4$-$5.3 & 40.57  & 1.11  & 1.49 & 0\farcs48\,$\times$\,0\farcs34 & $-$83.9 & 45.9 \\
      & 3 & 06 Jan 2022 & 45 & 14.9$-$976.6 & 0.4$-$5.1 & 40.57  & 0.71  & 1.50 &  &  &  \\

      & 1 & 03 Jan 2022 & 42 & 14.9$-$783.1 & 0.4$-$5.3 & 40.55  & 1.27  & 1.37 &  &  &  \\
     3 & 2 & 07 Jan 2022 & 46 & 14.9$-$976.6 & 0.4$-$5.5 & 40.57  & 2.29  & 0.99 & 0\farcs48\,$\times$\,0\farcs34 & $-$83.2 & 57.1 \\
      & 3 & 07 Jan 2022 & 45 & 14.9$-$976.6 & 0.4$-$5.4 & 40.58  & 1.79  & 1.27 &  &  &  \\

      & 1 & 07 Jan 2022 & 45 & 14.9$-$976.6 & 0.4$-$5.4 & 40.58  & 1.55  & 1.29 &  &  &  \\
     4 & 2 & 08 Jan 2022 & 44 & 14.9$-$976.6 & 0.4$-$5.4 & 40.57  & 2.39  & 0.76 &         0\farcs83\,$\times$\,0\farcs53 & $-$77.4 & 40.3 \\
      & 3 & 13 Apr 2022 & 45 & 15.0$-$500.2 & 0.8$-$8.5 & 40.57  & 0.38  & 1.50 &  &  &  \\
      & 4 & 13 Apr 2022 & 46 & 15.0$-$500.2 & 0.8$-$8.5 & 40.55  & 0.43  & 1.50 &  &  &  \\

      & 1 & 14 Apr 2022 & 44 & 15.0$-$500.2 & 0.7$-$8.2 & 40.55  & 0.43  & 1.50 &  &  &  \\
     5 & 2 & 15 Apr 2022 & 44 & 15.0$-$500.2 & 0.7$-$7.5 & 40.55  & 0.88  & 1.50 & 0\farcs85\,$\times$\,0\farcs63 & 88.3 & 48.2 \\
      & 3 & 17 Apr 2022 & 46 & 15.0$-$500.2 & 0.7$-$7.2 & 40.57  & 0.50  & 1.50 &  &  &  \\

      & 1 & 14 Apr 2022 & 44 & 15.0$-$500.2 & 0.7$-$8.2 & 40.55  & 0.57  & 1.50 &  &  &  \\
     6 & 2 & 14 Apr 2022 & 44 & 15.0$-$500.2 & 0.7$-$8.2 & 40.57  & 0.34  & 1.50 & 0\farcs94\,$\times$\,0\farcs64 & $-$84.7 & 44.7 \\
      & 3 & 15 Apr 2022 & 42 & 15.0$-$500.2 & 0.7$-$7.7 & 40.55  & 0.52  & 1.50 &  &  &  \\
    \bottomrule
\end{tabular}
\end{table*}

\end{appendix}

\end{document}